\documentclass{aa}
\usepackage[varg]{txfonts}
\usepackage{natbib}
\usepackage{graphicx}
\graphicspath{{Images/}}
\usepackage{hyperref}
\hypersetup{
    colorlinks=false, 
    linkcolor=blue, 
    urlcolor=red, 
    linktoc=all 
}
\bibpunct{(}{)}{;}{a}{}{,}
\usepackage{xcolor}

\newcommand{\Halpha}{\ifmmode {\rm H}\alpha \else H$\alpha$\fi}
\newcommand{\Hbeta}{\ifmmode {\rm H}\beta \else H$\beta$\fi}
\newcommand{\Hgamma}{\ifmmode {\rm H}\gamma \else H$\gamma$\fi}
\newcommand{\Hdelta}{\ifmmode {\rm H}\delta \else H$\delta$\fi}
\newcommand{\Lya}{\ifmmode {\rm Ly}\alpha \else Ly$\alpha$\fi}
\newcommand{\Lyb}{\ifmmode {\rm Ly}\beta \else Ly$\beta$\fi}
\newcommand{\HeI}{\ifmmode {\rm He}\,\textsc{i}\,\lambda5876 \else 
                  He\,\textsc{i}\,$\lambda5876$\fi}
\newcommand{\HeII}{\ifmmode {\rm He}\,\textsc{ii}\,\lambda4686 \else 
                   He\,\textsc{ii}\,$\lambda4686$\fi}
\newcommand{\hi}{H\,\textsc{i}}

\newcommand{\nev}{[Ne\,\textsc{v}]}
\newcommand{\Nevir}{[Ne\,\textsc{v}]$14.3,24.3\mu\rm{m}$}

\newcommand{\cii}{C\,\textsc{ii}]}
\newcommand{\ciii}{\ifmmode {\rm C}\,\textsc{iii}] \else C\,\textsc{iii}]\fi}
\newcommand{\civ}{\ifmmode {\rm C}\,\textsc{iv} \else C\,\textsc{iv}\fi}

\newcommand{\Oii}{[O\,\textsc{ii}]\,$\lambda\lambda 3726,3729$}
\newcommand{\oiii}{[O\,\textsc{iii}]}
\newcommand{\Oiii}{[O\,\textsc{iii}]\,$\lambda 5007$}

\newcommand{\mgii}{Mg\,\textsc{ii}}

\newcommand{\ebv}{$\rm{E}(\rm{B}-\rm{V})$}

\newcommand{\eddrat}{\ifmmode{\lambda_{\text{Edd}}} \else  $\lambda_{\text{Edd}}$ \fi}
\newcommand{\mbh}{\ifmmode{M_{\text{BH}}} \else  $M_{\text{BH}}$ \fi}
\begin{document}

\title{COSMOS2020: Investigating the AGN-obscured accretion phase at $z\sim1$ via \nev\ selection}

\author{Luigi Barchiesi\thanks{\email{luigi.barchiesi2@unibo.it}}\inst{1,2}
  \and C.~Vignali\inst{1,2}
    \and F.~Pozzi\inst{1,2}
        \and R.~Gilli\inst{2}
        \and M.~Mignoli\inst{2}
        \and C.~Gruppioni\inst{2}
        \and A.~Lapi\inst{3}
        \and S. Marchesi\inst{1,2,4}
        \and F.~Ricci\inst{5,2}
        \and C.~M.~Urry\inst{6}}


\institute{Dipartimento di Fisica e Astronomia ``Augusto Righi'', Universit\`a degli Studi di Bologna, via P. Gobetti 93/2, 40129 Bologna, Italy
  \and INAF-OAS, Osservatorio di Astrofisica e Scienza dello Spazio di Bologna, via Gobetti 93/3, 40129 Bologna, Italy
  \and SISSA, Via Bonomea 265, 34136 Trieste, Italy
  \and Department of Physics and Astronomy, Clemson University, Kinard Lab of Physics, Clemson, SC 29634, USA
  \and Dipartimento di Matematica e Fisica, Università Roma Tre, via della Vasca Navale 84, I-00146, Roma, Italy
  \and Department of Physics and Yale Center for Astronomy \& Astrophysics, Yale University, New Haven, CT 06520, USA}


\date{Accepted 14 February 2024}

\abstract{The black hole-and-galaxy (BH-galaxy) co-evolution paradigm predicts a phase where most of the star formation (SF) and BH accretion takes place in gas-rich environments, namely, in what are likely to be very obscured conditions. In the first phase of this growth, some of the galactic gas is funnelled toward the centre of the galaxy and is accreted into the supermassive BH, triggering  active galactic nucleus (AGN) activity. The large quantity of gas and dust hides the emission and the AGN appears as an obscured (type 2) AGN. The degree of obscuration in type 2 AGNs may even reach values as high as $N_H>10^{24}$ cm$^{-2}$ (i.e., Compton-thick, CT). Population synthesis models of the X-ray background (XRB) suggest that a large population of CT-AGN is, in fact, needed to explain the still unresolved XRB emission at energy above 20 keV. 

In this work, we investigated the properties of 94 \nev$3426\AA$-selected type 2 AGN in COSMOS at $z=0.6-1.2$, performing optical-to-far-infrared (FIR) spectral energy distribution (SED) fitting of COSMOS2020 photometric data to estimate the AGN bolometric luminosity and stellar mass, star formation rate, age of the oldest stars, and molecular gas mass for their host-galaxy. In addition, we performed an X-ray spectral analysis of the 36 X-ray-detected sources to obtain reliable values of the AGN obscuration and intrinsic luminosity, as well as to constrain the AGN properties of the X-ray-undetected sources.

We found that more than two-thirds of our sample is composed of very obscured sources ($\rm{N}_{\rm{H}}>10^{23}$ cm$^{-2}$), with about 20\% of the sources being candidate CT-AGN and half being AGNs in a strong phase of accretion ($\lambda_{\text{Edd}} > 0.1$). We built a mass- and redshift-matched control sample and its comparison with the \nev{} sample indicates that the latter has a higher fraction of sources within the main sequence of star-forming galaxies and shows little evidence for AGNs quenching the SF. As the two samples have similar amounts of cold gas available to fuel the SF, this difference points towards a higher efficiency in forming stars in the \nev-selected sample. The comparison with the prediction from the in situ co-evolution model suggests that \nev{} is an effective tool for selecting galaxies in the obscured growth phase of the BH-galaxy co-evolution paradigm. We find that the "quenching phase" is still to come for most of the sample and only few galaxies show evidence of quenched SF activity.
}

\maketitle

\section{Introduction}

The study of obscured (type 2) active galactic nuclei (AGNs) is fundamental to understanding the evolution phases of AGNs and their influence on galaxy growth. The presence of scaling relations between the mass of the supermassive black hole (SMBH) and several physical properties of the host galaxy, such as galaxy bulge mass, luminosity, and velocity dispersion (e.g., \citealt{kormendy95,magorrian98,ferrarese00,gebhardt00,kormendy13}), suggest that the growth of the SMBH and of the galaxy are coupled, leading to the formulation of the AGN-galaxy co-evolution paradigm (e.g., \citealt{hopkins08,lapi14,lapi18}). In this scenario, an intense phase of star formation (SF) is triggered by a wet merger, at least for the most luminous and massive systems \citep{silk98,dimatteo05,treister12,lamastra13}, or by in situ processes including fast inflow and cooling of gas clumps (e.g., \citealt{lapi18}). A fraction of the gas reservoir of the galaxy is funnelled towards the SMBH and initiates AGN activity. Thus, this phase is characterized by the growth of both the SMBH and the stellar mass of the galaxy. Due to the large quantity of gas, the majority of the AGN-emitted radiation is absorbed and the source appears as an obscured AGN. We can find indications of this phase in the larger fraction of obscured AGNs found in mergers and post-mergers galaxies \citep[e.g.,][]{satyapal14,ellison19,secrest20}, as well as in the detection of AGN activity in strongly star-forming submillimeter galaxies at high-$z$ \citep[e.g.,][]{archibald02,almaini03,alexander05}.
As the SF and the AGN consume, heat up, and expel the gas, the AGN begins to appear less obscured and is identified with a type 1 AGN. Moreover, the decrease in the amount of available gas leads to a decrease of the SF and, eventually, to the end of the AGN activity, leaving behind a ``red-and-dead'' elliptical galaxy (e.g., \citealt{hopkins08,cattaneo09}). This scenario is supported by the observations of molecular outflows, extending on kpc scales from the nucleus, in some AGN hosts and ultra-luminous infrared galaxies (ULIRGs) (e.g., \citealt{feruglio10,glikman12,cicone14}). In these systems, the mass loss ranges from one to several times the star formation rate (SFR).

Similarly, using various emission line diagnostics, neutral (\hi\ 21 cm, \cii\ $158\mu$m) and ionized (\oiii\ $5007\AA$, \Halpha\ $6564\AA$) outflows were also detected (e.g., \citealt{nesvadba08,alexander10,harrison12}, see \citealt{harrison18} for a review). Ionized AGN outflows tend to be more common ($\sim 75\%$, e.g., \citealt{brusa15,brusa16,perna15a,perna15b,kakkad16,zakamska16,lamassa17,toba17} in high luminosity and high Eddington ratios and in red and obscured dusty sources, as expected from the evolutionary model by \citet{hopkins07}. Studying the obscured phase and how the AGN gets rid of the surrounding material is crucial to testing and improving our understanding of the AGN-galaxy co-evolution scenario.\par
Furthermore, AGNs are the main contributors to the X-ray background (XRB). The integrated emission of the AGNs, obtained from the X-ray deepest surveys, can account for most of the XRB surface brightness below 10 keV \citep{xue11,moretti12}. However, where the XRB spectra peaks ($20-40$ keV) most of the XRB is still unresolved. Extrapolating at these energies the spectrum from the deep surveys showed that an additional large population of heavily obscured, Compton-thick (CT) AGNs with N$_{\text{H}}>10^{24}$ cm$^{-2}$ is required to fully reproduce the XRB spectra. Furthermore, XRB population synthesis models predict that the maximum contribution to the XRB of the ``missing'' CT objects in the $L_{x}^{intr} \approx 10^{42}-10^{44}$ erg/s luminosity range should peak at $z \approx 1$ \citep{gilli07, ueda14, ananna19}. As it is difficult to constrain the abundance and the properties of these CT AGNs using only XRB synthesis models, studies of individually detected sources are needed \citep{gilli13}.\par
Although several methods for selecting type 2 AGNs are available (see \citealt{vignali14_osc} for a review), a complete census of these objects cannot be achieved using only a single observing band. The X-ray radiation, originated by the hot corona in the innermost region of the AGN, is a good tracer of the AGN intrinsic emission, however, when the nucleus is obscured by column densities as large as $\sim 10^{24-25} \text{cm}^{-2}$, even hard X-rays are severely depressed. Mid-infrared wavelengths can be effectively used as an X-ray complementary selection method, as the optical-to-X-ray absorbed radiation is re-emitted at these wavelengths after being thermally reprocessed by the obscuring torus. However, distinguishing between the AGN and the SF contributions to the total infrared (IR) emission is not trivial, especially for galaxies with high SFR and strong PAH features. Future planned mid-IR cryogenic observatories (e.g., the PRobe far-Infrared Mission for Astrophysics, PRIMA,  \citealt{bradford22}), with higher sensitivity and better spectral resolution than what is currently available, will have the capabilities to both detect these sources up to very high-redshift and to distinguish the AGN and SF contribution \citep{spinoglio21, barchiesi21_spica, bisigello21a}. Finally, obscured sources can also be identified using mm-observations, as they can be sensible up to $N_H \sim 10^{26}\,\rm{cm}^{-2}$ (e.g., \citealt{behar15,kawamuro22}). 
\par
Type 2 AGNs can be also selected using narrow, high-ionization emission lines, such as the \oiii$5007\AA$, the \nev$3426\AA$ and the  \civ$1549\AA$ narrow emission lines. These lines, produced in the narrow line region (NLR), do not suffer from the nuclear extinction and their flux is a better proxy of the AGN intrinsic emission. In particular, the \nev\ line, despite being $\sim 9$ times fainter than the \oiii\ line, allows for AGN selection up to $z\approx 1.5$, whereas the \oiii\ is redshifted out of the optical range at $z\approx 0.8$. Moreover, due to its high-ionization potential of 97 eV (vs. 54 eV of \oiii), it is an unambiguous marker of AGN activity (e.g., \citealt{gilli10,mignoli13,cleri22}).
Several works have shown that pairing AGN optical lines selection with X-ray data is an effective method to find obscured and CT AGNs (e.g., \citealt{ maiolino98,cappi06,vignali06,vignali10,gilli10,mignoli13,mignoli19}). Among those, \citet{gilli10} analyzed a sample of 74 local objects with \nev\ and X-ray detection. They showed that the \nev\ line is a good tracer of AGNs and that the selection of type 2 AGNs via this line provides a similar fraction of obscured AGN as the \oiii{} selection. They also show that the use of the \nev{} line should not be biased toward high-luminosity AGNs, but it could be toward low NLR extinction, as even a modest value $\rm(E(B-V)=0.5$ would dump the flux by a factor of 10. \citet{gruppioni16} analyzed a sample of 76 Seyfert  galaxies and found that the \Nevir\ emission lines are good tracers of the AGN intrinsic power and calibrated two relations between the luminosity of these lines and the AGN bolometric luminosity.\par
The use of different emission lines to select AGNs at various redshifts (i.e., $z \lesssim 0.8$ for \oiii, $0.6 \lesssim z \lesssim 1.5$ for \nev\ and $1.5 \lesssim z \lesssim 3$ for \civ) allows for the study of the redshift evolution of both the AGN and the host properties \citep{vignali10,gilli10,mignoli13,vignali14,mignoli19}.\par
The X/\nev\ flux ratio was used by \citet{vignali14}, hereafter V14, to trace the obscuration of a sample of \nev -selected type 2 AGN in the C-COSMOS field. In fact, both the observed X-ray and the \nev\ fluxes are linked to the intrinsic AGN emission, but the X-ray flux also suffers from the source obscuration. \citet{gilli10} calibrated a relation between X/\nev\ and the N$_{\text{H}}$ using a sample of 74 bright, nearby Seyferts with both X-ray and \nev\ data and for which the column density was determined unambiguously. They found that  the mean X/\nev\ ratio for unobscured Seyferts is about 400, about 80\% of local Seyferts with X/\nev$<100$ are obscured by column densities above $10^{23}$ cm$^{-2}$ and essentially all objects with observed X/\nev$<15$ are CT (see also \citealt{cleri22}).\par
In this paper, we present the properties of a sample of 94 \nev-selected type 2 AGNs, obtained via X-ray spectral analysis and from optical-to-FIR spectral energy distribution (SED) fitting.  The use of the \nev\ selection method restricted the sample to AGNs in the $0.65< z <1.2 $ redshift range, where most of the XRB ``missing'' sources are expected to lie \citep{gilli13}. This work is an extension of previous works (\citealt{mignoli13}, V14) for the AGNs in the C-COSMOS field. We made use of newer X-ray data from the Chandra COSMOS Legacy catalog \citep{civano16,marchesi16b}, which extend the X-ray coverage of the COSMOS field from 0.9  to 2.2 deg$^2$ and provide a more uniform coverage. We also studied the hosts of type 2 AGNs to characterize their parameters, stellar mass, and star formation rate. We investigated whether these galaxies are different from ``normal'' galaxies due to the AGN influence. The optical selection of the sample is presented in \S~\ref{sec:sample}, with the X-ray data in \S~\ref{sec:sample_x} and the photometric data in \S~\ref{sec:sample_ott}. The X-ray analysis results are reported in \S~\ref{sec:x-ray}, along with the fraction of CT objects. The SED-fitting algorithm and its results are presented in \S~\ref{sec:sed_fitting}. In \S~\ref{sec:parent} we present the comparison with a stellar mass- and redshift-matched control sample of non-active galaxies and the interpretation of our results in the light of the in situ co-evolution scenario is in \S~\ref{sec:insitu}. Our conclusions are then reported in \S~\ref{sec:conclusion}. Throughout this paper, we adopt the following cosmological parameters: H$_0 = 70\, \text{km}\, \text{s}^{-1}\, \text{Mpc}^{-1}$, $\Omega_{\text{M}} = 0.3,$ and $\Omega_{\Lambda} = 0.7$ \citep{spergel03}.

\section{Sample}\label{sec:sample}
We studied the \nev\ type 2 AGN sample described in \citet[][hereafter M13]{mignoli13}. It was derived from the zCOSMOS-Bright spectroscopic survey \citep{lilly07, lilly09}, which provided the 5500 – 9700 $\AA$ spectra of $\sim 20000$ objects in the COSMOS \citep{scoville07} field. From the zCOSMOS-Bright catalog, we selected 94 type 2 AGNs in the redshift range  of $\sim 0.65-1.20$ on the basis of \nev\ detections and their spectral properties. The redshift range assured that both the \nev3346\AA\ and the \nev3426\AA\ emission line fall within the spectral coverage. Sources previously identified as type 1 AGNs due to broad (>1000 km s$^{-1}$) emission lines in their spectra were excluded. The sample was composed of sources from redshift $z=0.6606 $ to $z=1.1767$, with mean $z=0.85 \pm 0.13$ and a median $z=0.86$. The mean (aperture corrected) \nev\  flux was $F_{\text{\nev}}= (1.81 \pm 1.23) \times 10^{-17} $ erg cm$^{-2}$ s$^{-1}$, with median $F_{\text{\nev}}= 1.44  \times 10^{-17} $ erg cm$^{-2}$ s$^{-1}$; the mean \nev\ equivalent width is EW$_{\text{\nev}}= 18.17 \pm 15.81$ \AA\ with median EW$_{\text{\nev}}= 13.9$ \AA\ (see M13 for further details).\par
Due to the complex selection of the \nev\ sample, judging its completeness is not trivial, although M13 tried to draw some conclusions based on the comparison with other selection techniques. They found that the selection via \nev\ emission line is complementary to the one based on the lack of broad emission lines in the optical spectra coupled with a $L_{\rm{2-10keV}}>10^{42}\,\rm{erg\,s^{-1}}$. On the one hand, a significant fraction ($\sim60\,\% $) of both X-ray- and \nev-selected AGNs would not be classified as AGN via the so-called ``blue diagnostic diagram'' (i.e., exploiting the \Oiii, \Hbeta, and \Oii\ line ratios), probably due to the fact that in these objects AGN and SF coexist. On the other hand, only $\sim10\,\%$ ($\sim9\,\%$) of the ``blue diagnostic diagram''-selected AGN have \nev\ (X-ray) detection, suggesting the complementarity of these selection techniques. The majority ($\sim80\,\%$) of the \nev-selected AGNs are also AGNs according to the mass-excitation (MEx) diagram of \citet{juneau11}. The fact that the other selection techniques find a sizable number of AGNs without \nev\ detection could be linked to obscuration in the host galaxy. In fact, as shown by \citet{gilli13}, even a modest \ebv\ is able to extinct the \nev\ emission.  Finally, the M13 analysis of the VIMOS spectra of the \nev\ sources showed low \ebv, suggesting again that this selection may lose some of the \nev\ emitter AGNs due to extinction in their host-galaxy.\par
The redshift distribution of the \nev\ AGNs clearly follows the one of its parent sample \citep[the zCOSMOS galaxies; see Fig.1 of][]{mignoli13} due to the magnitude-limited survey selecting equally well host-galaxy dominated type 2 AGNs and normal galaxies. The comparison with other AGN samples from the literature can be found at the end of \S~\ref{cap_x_detec}.

\subsection{X-ray Data}\label{sec:sample_x}
All 94 \nev-selected sources fall in the \textit{Chandra}-COSMOS Legacy mosaic, which is composed of data from the C-COSMOS survey \citep[the central $\sim0.9\,$deg$^2$;][]{elvis09} and from the COSMOS Legacy survey \citep[covering the external $\sim 1.7$ deg$^2$ with a similar depth of the C-COSMOS survey;][]{civano16}. The whole mosaic covered $\sim 2.2$ deg$^2$ with a total exposure time of $\sim 4.6$ Ms. Thanks to the broader coverage of the COSMOS Legacy survey, we had 23 more sources with X-ray coverage with respect to V14; moreover, we included two sources which were previously excluded because they fell in bad positions of the C-COSMOS mosaic. \\ Following \citet{vignali14}, we found 36 \nev-selected type 2 AGNs detected by \textit{Chandra} within 1.3" from the optical position,  with a median displacement of 0.41". We visually inspected the 7 sources with counterparts farther than 0.75" and found that they are good matches. The X-ray spectra were extracted as described in \citet{marchesi16b}, using the \textit{CIAO} \citep{ciao} tool \texttt{specextract} from circular regions of radius $r_{90}$ (i.e., the radius that contains 90\% of the PSF in the 0.5 - 7 keV observed-frame band). The background spectra were extracted from annuli centered on the source position with inner radius $r_{90}+2.5$" and outer radius of $r_{90} + 20$", paying attention to avoid the inclusion of X-ray sources. For each source, the spectra of its observations were combined in a single spectrum via the \textit{CIAO} tool \texttt{combine\_spectra}.

\subsection{Photometric data}\label{sec:sample_ott}
The optical and IR data used to identify the \nev\ sources were taken from the COSMOS 2020 catalog \citep{cosmos2020} (improved version of the previous COSMOS 2015 catalog \citealt{laigle16}), which contains photometry in 44 bands (from 1526\AA\ to 8$\mu$m) for $\sim 1.7$ million objects in the 2 deg$^2$ COSMOS field, along with matches with X-ray, near-ultraviolet (near-UV), and far-IR data. We used 3" aperture (corresponding to a physical size of $23.4\,\rm{kpc}$ at the median redshift of our sample) fluxes from 20 photometric bands and the COSMOS2020 matches with the $24\mu$m band from the MIPS (Multi-Band Imaging Photometer) detector onboard \textit{Spitzer}, with the $100\,\mu$m and $160\mu$m bands and the $250\,\mu$m, $350\,\mu$m, and $500\,\mu$m bands from the PACS and SPIRE detectors of \textit{Herschel}, and with the $850\,\mu$m from the SCUBA instrument at \textit{JCMT} (we refer to \citealt{cosmos2020} for a complete description of the COSMOS2020 catalog and source associations). In total, we exploited data from 31 filters (see Table~\ref{tab:laigle_band}).\par
Two sources of the NeV sample were not in the latest COSMOS 2020 catalog (at the time of writing); for these sources, we used the photometric information from the COSMOS 2015 catalog. As we analyzed the whole sample also with the COSMOS 2015 catalog and we found no systematic difference between the values obtained using one or the other catalog, we are confident in reporting the results of these two sources along with those obtained with the newest COSMOS 2020 catalog. 

\begin{table}
\caption{Summary of COSMOS2020 photometric bands used in this work.  The effective wavelength is the median wavelength weighted by transmission and the widths are defined as the difference between the maximum and the minimum wavelengths (calculated as the first and the last wavelengths with a transmission of at least 1\%).}
\label{tab:laigle_band}
\centering
\begin{tabular}{cccc}
\hline \hline
Instrument & Filter & Effective & Width    \\
/Survey & & $\lambda$ [\AA] & [\AA]\\
\hline
MegaCam/CFHT & \textit{u*} & 3823.3 & 670 \\
\hline
Suprime-Cam     & IB427         &       4256.05         &       305.6 \\
/Subaru & B     &       4400.33 &       1399.1          \\
        &       IB464   &       4633.48 &       330.5   \\
        &       IB505   &       5060.57 &       378.1   \\
        &       IB527   &       5261.1          &       242     \\
        &       V       &       5477.8  &       955     \\
        &       IB574   &       5764.8          &       271.5   \\
        &       $r$     &       6136.24 &       1918    \\
        &       IB679   &       6778.75 &       555.3   \\
        &       IB709   &       7070.67 &       511.5   \\
        &       IB738   &       7358.64 &       490.2   \\
        &       $i^+$   &       7630.05 &       1872.5  \\
        &       IB827   &       8241.69 &       514.3   \\
        &       $z^{++}$&       9020.18 &       1960.4  \\
        &       Y       &       9759.16 &       1752.8  \\
\hline                                                  
VIRCAM  &       Y$^{UD}$        &       10214.2 &       970     \\
/VISTA  &       J$^{UD}$        &       12534.6 &       1720    \\
(UltraVISTA-DR2)        &       H$^{UD}$        &       16453.4 &       2900    \\
        &       K$^{UD}_{S}$    &       21539.9 &       3090    \\
\hline                                                  
IRAC/\textit{Spitzer}   &       ch1     &       35634.3 &       7460    \\
(SPLASH)        &       ch2     &       45110.1 &       10110   \\
        &       ch3     &       57593.4 &       14140   \\
        &       ch4     &       79594.9 &       28760   \\      
\hline
MIPS/\textit{Spitzer} & $24\mu$m        &       232096  &       110494  \\
\hline
PACS/\textit{Herschel}& green           &       979036  &       558974          \\
        &       red     &       $1.54\times10^6$&       $1.26\times10^6$        \\
SPIRE/\textit{Herschel}&        PSW     &       $2.43\times10^6$        &       $1.26\times10^6$        \\
        &       PMW     &       $3.41\times10^6$        &       $1.46\times10^6$        \\
        &       PLW     &       $4.82\times10^6$        &       $2.92\times10^6$        \\              
\hline
SCUBA/\textit{JCMT}     &       $2.450\,$GHz    &       $4.48\times10^6$&       $1.04\times10^6$        \\      
\hline
\end{tabular}
\end{table}

\section{X-ray spectral analysis}\label{sec:x-ray}

\subsection{X-ray-detected sources}\label{cap_x_detec}
Using the \textit{XSPEC} software \citep{xspec} we performed the X-ray spectral analysis of 36 type 2 AGNs with X-ray detection. The median number of net (i.e., background-subtracted) counts is 85, four sources have 15 counts or less. We divided the 36 sources into two sub-samples, on the basis of their net counts. The high-counts sample is composed of 17 sources with at least 90 net counts, and the low-counts sample of 19 sources with less than 90 net counts. For the low-counts sample, we used unbinned data and \textit{C}-statistic \citep{cash79}; for the high-counts sample we rebinned the data at 25, 15, and 10 counts per bin (respectively, for sources with net-counts $>500$, $>200,$ and $>100$) and used Gaussian statistics.\par
We refer to Appendix~\ref{sec:xray_prop} for the description of the X-ray models used in the spectral fitting, as well as for the complete spectral properties of all the sources. We first fit the sources with a power-law model, modified by the Galactic absorption obtaining a mean photon index value of $\Gamma=0.82$ with a standard deviation of $0.90$. We find that 6 sources have $\Gamma \geq 1.6$, typical of unobscured AGNs, while 16 objects have $\Gamma \leq 1.0$; of these 7 have negative spectral index, usually found in very obscured sources.\par
To characterize the sources in terms of obscuration, we fixed the photon index to a value of $\Gamma=1.8$, typical of unobscured sources, and added an absorption component to model the source obscuration. The mean N$_{\text{H}}$ of the high-count sample is $\approx 7.5 \times 10^{22}$ cm$^{-2}$, while for the low-count sample we obtained $\approx 32.2 \times 10^{22}$ cm$^{-2}$. On the basis of their obscuration, we classified nine sources as highly obscured (N$_{\text{H}} > 10^{23}$ cm$^{-2}$) and two as CT objects. Twelve sources have low values of obscuration ($N_{\text{H}}<10^{22}$ cm$^{-2}$).
We computed the intrinsic (i.e., absorption-corrected) $2-10$ keV rest-frame luminosity of the sample, obtaining a mean value of $8.3 \times 10^{43}$ erg/s for the high-counts sample and $2.7 \times 10^{43}$ erg/s for the low counts sample. The median (and $16^{\rm{th}}-84^{\rm{th}}$ percentiles) intrinsic $2-10$ keV rest-frame luminosity of the whole sample is $\log{(L_{\rm{2-10keV}}/\rm{erg\,s^{-1}})}=43.6_{-0.4}^{+0.6}$.\par
As the sources have low numbers of counts we preferred using simple phenomenological models. However, few sources display more complex spectra that call for deeper investigations. Source \textit{lid}1840 (COSMOS Legacy ID) with 238 net-counts and N$_{\text{H}}<0.5 \times 10^{22}\, \text{cm}^{-2} $ shows the possible presence of a 6.9 keV line with a significance of $\approx 2.6\, \sigma$ and an equivalent width of EW$_{6.9}=0.51 \pm 0.38$ keV. Source \textit{cid}1508 (C-COSMOS ID) has a spectrum that shows the presence of a possible soft excess with respect to the absorbed power law model. We substituted the absorption component with a partial covering fraction absorption (\texttt{zpcfabs}) and found a lower limit on the covering fraction $f>0.77$, while the value of N$_{\text{H}}$ being compatible with the one obtained using the simple absorption component. For source \textit{cid}138 (99 net-counts) we found a variation of the net-count rate of more than a factor of 6 (with a significance of $6.7\sigma$) between the 2007 and 2014 observations. We did not find any significant variation in the hardness ratio, spectral index or N$_{\text{H}}$. As the flux variation is not accompanied by a variation of the spectral properties, we hypothesized it may be linked to differences in the AGN accretion rate.\par
\begin{figure}
  \centering
  \resizebox{\hsize}{!}{\includegraphics{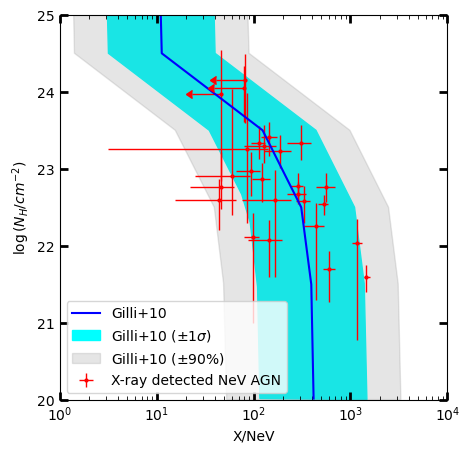}}
  \caption{X/\nev\ vs N$_{\text{H}}$ diagram. Red dots are the X-ray-detected sources of the \nev\ sample, with N$_{\text{H}}$ obtained from the X-ray spectral analysis. The solid line shows the expected X/\nev\ values as a function of absorption, as computed by \citet{gilli10} (but see also \citealt{cleri22}) using spectral templates with different N$_{\text{H}}$, starting from the mean X/\nev\ obtained from a sample of unobscured Seyfert galaxies. The cyan shaded region was computed in the same way, but starting from the mean X/\nev\ $\pm 1 \sigma$ and the grey shaded region starting from the mean X/\nev\ $\pm 90\%$. The \nev\ sample populates the obscured AGN region of the diagram, i.e., $N_{H}\gtrsim 10^{22}$ cm$^{-2}$.}
  \label{fig:autostrada} 
\end{figure}

We computed the 2-10 keV rest-frame flux (not corrected for obscuration) of the sample, obtaining a median flux of F$_{2-10}=0.61 \times 10^{-14}$ erg/s/cm$^2$. We compared the fluxes with those from V14, which were obtained via an \textit{Xspec} spectral fitting with a power-law model with $\Gamma=1.4$. Except for \textit{cid}138 the values are in agreement within their errors. We used the 2-10 keV rest-frame flux and the \nev\  flux from M13 to compute the X/\nev\  ratio. The mean X/\nev\ is 313 with a standard deviation of 321. Ten sources had X/\nev\ $<100$. No source has X/\nev\ lower than 15. In Fig.~\ref{fig:autostrada}, we compared our data with the X/\nev\ vs N$_{\text{H}}$ diagram obtained by \citet{gilli10}.
The plot was produced using the spectral templates of \citet{gilli07}. These are AGN X-ray spectral models with a primary power-law with $\Gamma=1.9$, cut-off energy E$_C=200$ keV, a variety of absorptions (log N$_{\text{H}})=21.5,22.5,23.5,24.5,>25$, a 6.4 keV emission line and, in case of obscured spectra, a 3\% soft scattered component. The blue solid line was obtained using the mean X/\nev\ ratio of a sample of 74 unobscured Seyfert galaxies in the local Universe and, starting from it, computing the expected X/\nev\  ratio at increasing levels of absorption, using the spectral templates. The same computation was carried out starting from the mean X/\nev\  ratio $\pm 1 \sigma$ and $\pm 90\%$, to produce the $ 1 \sigma$ and $ 90\%$ limits. The procedure is extensively described in \citet{gilli10}.
As we can see from Fig.~\ref{fig:autostrada}, the X-ray-detected sources of the \nev\ sample populate the obscured quasar region of the diagram, i.e., $10^{22} < \text{N}_{\text{H}}<10^{24}$ cm$^{-2}$, with a few sources in the unobscured region ($\text{N}_{\text{H}}<10^{22}$ cm$^{-2}$); the majority of the sources lie within the 1 $\sigma$ limit. \par

We compared the results of our spectral analysis with those presented by \citet{marchesi16b}, hereafter M16, and \citet[who studied CT-candidate from M16]{lanzuisi18}. M16 performed an X-ray spectral analysis of the $\sim 1850$ extragalactic sources in the \textit{Chandra} COSMOS-Legacy survey with more than 30 net-counts in the $0.5-7\,$keV band. Spectra were fitted with a fixed photon-index $\Gamma=1.9$ power-law, an absorption component, an optional un-absorbed second power-law to model the scattered emission, and a $6.4\,$keV (rest frame) Gaussian to reproduce the iron K$\alpha$ emission line. Out of our 36 X-ray-detected sources in our sample, 6 of them are not in the M16 work, as they are below the 20 net-counts threshold. For the remaining 30 sources, we found an excellent agreement in $\text{L}^{\text{intr}}_{2-10\text{keV}}$ and $\text{N}_{\text{H}}$: the median ($16^{\rm{th}}$ and $84^{\rm{th}}$ percentile) ratio between M16 and ours $\text{L}^{\text{intr}}_{2-10\text{keV}}$ and $\text{N}_{\text{H}}$ are $1.000_{-0.003}^{+0.004}$ and $1.02_{-0.02}^{+0.10}$, respectively. Regarding the two CT X-ray-detected sources, \textit{cid}1019 is classified as CT in \citet{lanzuisi18} as well, while \textit{cid}1706 has 15 net counts and was not included in M16; the other 5 sources that are not in M16 are all obscured AGNs, 2 of them with $\text{N}_{\text{H}}>10^{23}\,\text{cm}^{-2}$.\par
Compared to the sub-sample of type 2 AGNs of M16, which is composed of 1111 sources at $0.066<z<4.45$, the current sample covers a narrower redshift range and is characterized by a lower average redshift ($z=0.85$ vs $z=1.05$ of M16). As we did not exclude the sources with less than 30 net counts, we found a lower average number of net counts (85 vs 102) and, on average, slightly less luminous AGNs with mean ($\pm \sigma$) $\log{(L^{\nev}_{\rm{2-10keV}}/\rm{erg\,s^{-1}})}=43.5\pm 0.5$ vs $\log{(L^{\rm{M16}}_{\rm{2-10keV}}/\rm{erg\,s^{-1}})}=43.7 \pm 0.6$. Finally, we note that our sample is on average as obscured as the one from M16 ($\log{(N_{\rm{H}}^{\nev}/\rm{cm^{-2}})}=22.8_{-0.3}^{+0.6}$ vs $\log{(N_{\rm{H}}^{\rm{M16}}/\rm{cm^{-2})}=22.8\pm0.3}$), but the inclusion of AGNs with less than 30 net counts highlights a tail of very obscured sources.\par
We also compared our X-ray-detected sample with the results of \citet[][hereafter I20]{iwasawa20}, who studied a sample of 185 bright sources detected in the \textit{XMM-Newton} deep survey ($3\,$Ms) of the \textit{Chandra} Deep Field South. In particular, we focused on the I20 lowz subsample, which encompassed 56 AGNs in the $0.4<z<1.0$ redshift range, with a median $z=0.7$ (thus slightly lower than our \nev\ sample). Their sample has on average less luminous and less obscured AGNs ($\log{(L^{\rm{I20}}_{\rm{2-10keV}}/\rm{erg\,s^{-1}})}=43.1\pm0.07$, $\log{(N^{\rm{I20}}_{\rm{H}}/\rm{cm^{-2}})}=22.0_{-0.1}^{+0.2}$)  than ours.\par

We selected a subsample of 8 \nev\ AGNs with the highest number of counts, sampling different amounts of obscuration. We fitted these sources with more realistic AGN models. We followed \citet{zhao21} and modeled the X-ray spectra with an absorbed intrinsic continuum, a reprocessed component (characterized by the \texttt{borus02} model), and a scattered component. We did not find any significant difference in intrinsic AGN luminosity and obscuration with respect to the values obtained with the simpler phenomenological model. We note, however, that the low-photon statistics forced us to fix the majority of the parameters (we chose to use the average values found by \citealt{zhao21} for a sample of very obscured AGNs with optimal X-ray spectral coverage): the photon index has been fixed at $\Gamma=1.8$, the cut-off energy at $E_{\rm{cut}}=500\,\rm{keV}$, the torus opening angle to $\theta_{\rm{Tor}}=48\deg$ (corresponding to a covering factor of $c_{\rm{f}}=0.67)$, the average column density to $N_{\rm{H,tor}}=1.24\times10^{24}\,\rm{cm^{-2}}$, and the fraction of scattered emission to $f_{\rm{scat}}=0.05$. Finally, we chose an inclination of $\theta_{\rm{obs}}=87\deg$, corresponding to an edge-on view of the sources.

\subsection{X-ray-undetected sources}\label{sec:xray_undetected}

Out of the 94 sources of the \nev\ sample, 58 have no X-ray detection. We ran a two-sample Kolmogorov–Smirnov (KS) test in order to investigate if X-ray-undetected sources are such because they fall in regions with shorter exposure-map derived time than those of the X-ray-detected sources. We built two empirical distribution functions of the exposure time for the X-ray-detected sources and for the X-ray-undetected, using the exposure maps derived from the entire COSMOS Legacy mosaic. We found, with a confidence of 89\% (KS statistic of $D=0.1197$), that the undetected sources are not associated to lower exposure times.\par

We used the \textit{CIAO} tool \texttt{scrflux} to calculate the net count-rate limit for each undetected source, and the tool \texttt{modelflux} to calculate the flux upper limit. Count rates were computed using all the observations for which the source falls in the field of view. Using \texttt{scrflux} we calculated the $0.5 - 7$ keV (observed-frame) net count-rate in a circular region centered on the source position, that contained 90\% of the PSF at 1 keV. Background counts were extracted in an annular region centered on the source position, using inner and outer radii of one and five times the radius of the source region. The uncertainties were computed at the 1$\sigma$ confidence level, exploiting the \citet{gehrels86}  approximation to confidence limits for Poisson distributions due to the low number of counts. Because the sources were not detected in the X-ray band, we can consider their net counts + $1\sigma$ uncertainties as the upper limits on the source net counts. \texttt{scrflux} computed the (upper limit) net count-rate by dividing the (upper limit) net counts by the effective (i.e., vignetted-corrected) exposure time at the source position. We used \texttt{modelflux} to calculate, from the upper limit net count-rates, the upper limit for the flux in the $2-10\,$keV rest-frame energy range. We used a power-law model with spectral index $\Gamma=0.4$ (the average spectral index of the low count sample), modified by Galactic absorption. This model takes into account the source obscuration via a flatter photon index than the intrinsic one $\Gamma=1.8-1.9$.\par
We used the rest-frame $2-10\,$keV flux upper limits, and the \nev\ fluxes to compute the X/\nev\ ratio upper limits. 54 sources have an upper limit $<100$, and 16 sources have X/\nev\  ratios $< 15$. This means that 93\% of the X-ray-undetected sources are candidate to be AGNs with N$_{\text{H}} > 10^{23}$ cm$^{-2}$, and 28\% to be CT AGNs. We used the X/\nev\ upper limits to compute lower limits on the N$_{\text{H}}$. Using the net-count rates, the N$_{\text{H}}$ lower limits, and a fixed $\Gamma=1.8$ spectral index, we estimate the intrinsic (i.e., absorption corrected) X-ray flux (and luminosity) upper limit via \texttt{modelflux}.\par
Testing this procedure on the X-ray-detected sources, we found a strong correlation between the intrinsic luminosities obtained in this way and those from the X-ray spectral analysis for the high-counts sample. The correlation is loose for the low-count sample, but for the majority of these sources, the two luminosities do not differ more than 2 $\sigma $. We did not find any particular trend between the accuracy of our method and the amount of obscuration, thus we concluded that it is a reliable way to estimate an upper limit on the X-ray intrinsic luminosity and that the major source of error is linked to the uncertainties of estimating the net-counts upper limit in case of low counts.\par
\begin{figure}
  \centering
  \resizebox{\hsize}{!}{\includegraphics{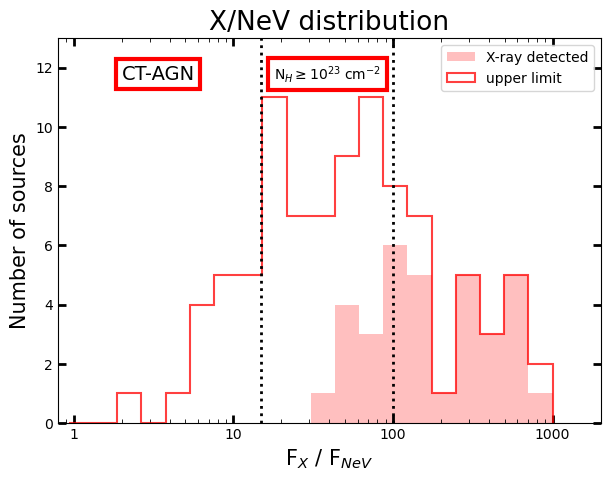}}
  \caption{X/\nev{} ratio distribution. Filled and empty histograms refer to X-ray detections and upper limits, respectively. The leftward area ($\text{X/\nev}\leq15$) is the region defined by \citet{gilli10} for Compton-thick AGNs, while the central area ($15<\text{X/\nev}\leq100$) is where very obscured (N$_{\text{H}} > 10^{23}$ cm$^{-2}$) and possible CT-AGNs should be located.}
  \label{fig:XneV_histo_all}
\end{figure}

  
Considering the whole (both X-ray-detected and undetected sources) \nev\ sample, at least 67\% of the sources have X/\nev\  ratios compatible with absorption N$_{\text{H}}> 10^{23}$ cm$^{-2}$, and at least 19\% of the sources are likely CT AGNs. The X/\nev\ flux ratios are plotted in Fig.~\ref{fig:XneV_histo_all}, the vertical dashed lines show the threshold,  defined by \citet{gilli10}, between Compton-thick (leftward direction) and Compton-thin (right-ward direction) sources. Although our sample is far from complete (see \S~\ref{sec:sample}), it is interesting to note that the fraction of CT-AGN is similar to the $f_{\rm{CT}}=0.24$ found by \citet{lanzuisi18} from a sample of 67 candidate CT-AGNs extracted from the M16 sample, although lower than the $f_{\rm{CT}}\approx 0.5$ expected from the distribution of the AGN population of \citet{buchenr15}.\par

\section{SED analysis}\label{sec:sed_fitting} 

We used the SED fitting technique to derive AGN and galaxy properties from photometric data. In Sect.~\ref{sec:sed_algorithm} we briefly present the SED-fitting algorithm as well as the AGN torus models we used. The results of the SED-fitting, along with the comparison with those obtained from the X-ray analysis, are presented in Sect.~\ref{sec:sed_results}.\par
\subsection{SED-fitting algorithm}\label{sec:sed_algorithm} 
We made use of the SED-fitting algorithm \textit{SED3FIT} \citep{berta13}, based on the \textit{MAGPHYS} \citep{cuna08} code, that performs SED-fitting with a combination of three components: stellar emission, dust emission from star formation and a possible dusty torus and AGN component. The stellar and dust emission are linked by energy balance arguments. Torus emission is independently included.\par
\textit{MAGPHYS} is a model package to interpret observed SEDs of galaxies (at rest wavelengths in the range $912 \AA\ < \lambda < 1 $mm) in terms of galaxy-wide physical parameters pertaining to the stars and the interstellar medium. The code uses two libraries of models: one that takes into account the stellar emission and the effects of dust attenuation, generated using the \citet{bruzual07} stellar population synthesis code with the attenuation computed using the angle-averaged model of \citet{charlot00}, the other that includes the IR emission of the dust, computed using the model of \citet{cuna08}. The optical and infrared libraries are linked together to provide the full SED of model galaxies from the far ultraviolet to the far-infrared wavelengths.\par
One of the main assumptions of the \textit{MAGPHYS} code is that the only significant source of dust heating is the starlight, thus ignoring any possible contribution of the AGN to the SED. The \textit{SED3FIT} code solves this limitation by adding a warm dust component to the modeled SED emission. It represents dust surrounding the active nucleus, assumed to be distributed in a toroidal region. The code uses $\chi^2$ minimization to find the best-fit model, by effectively fitting simultaneously the three components (stellar, dust, and AGN) to the data \citep[see][for further details]{berta13}.\par
The torus library we used to model the AGN contribution to the SED assumes that the AGN dust and gas are distributed in a toroidal shape, i.e., ``smooth-torus'' model. It was developed by \citet{fritz06} and updated by \citet{feltre12}. 
The geometry of the torus is a \textit{flared disc}. To reduce the calculation time, we selected only a sub-sample of 180 models between the $24\,000$ elements torus library. A detailed explanation of the torus models and parameters is reported in Appendix \ref{sec:torusmodels}. We instructed the SED-fitting code to run 100 different normalizations of the chosen torus models, for a total of $18\,000$ AGN spectra used in the SED-fitting procedure.\par
As a sanity check, we also performed additional runs of SED-fitting using the \texttt{X-CIGALE} SED-fitting code \citep{yang20,yang22}, the latest version of the code investigating galaxy emission (\texttt{CIGALE}; \citealt{burgarella05,noll09,boquien19}), which allowed us to use the newer \texttt{SKIRTOR} torus models \citep{stalevski12,stalevski16} and to model it as a ``clumpy'' medium. We did not find any significant difference in the AGN and host-galaxy properties.

\subsection{SED-fitting results}\label{sec:sed_results}
\begin{figure}
  \centering
  \resizebox{\hsize}{!}{\includegraphics{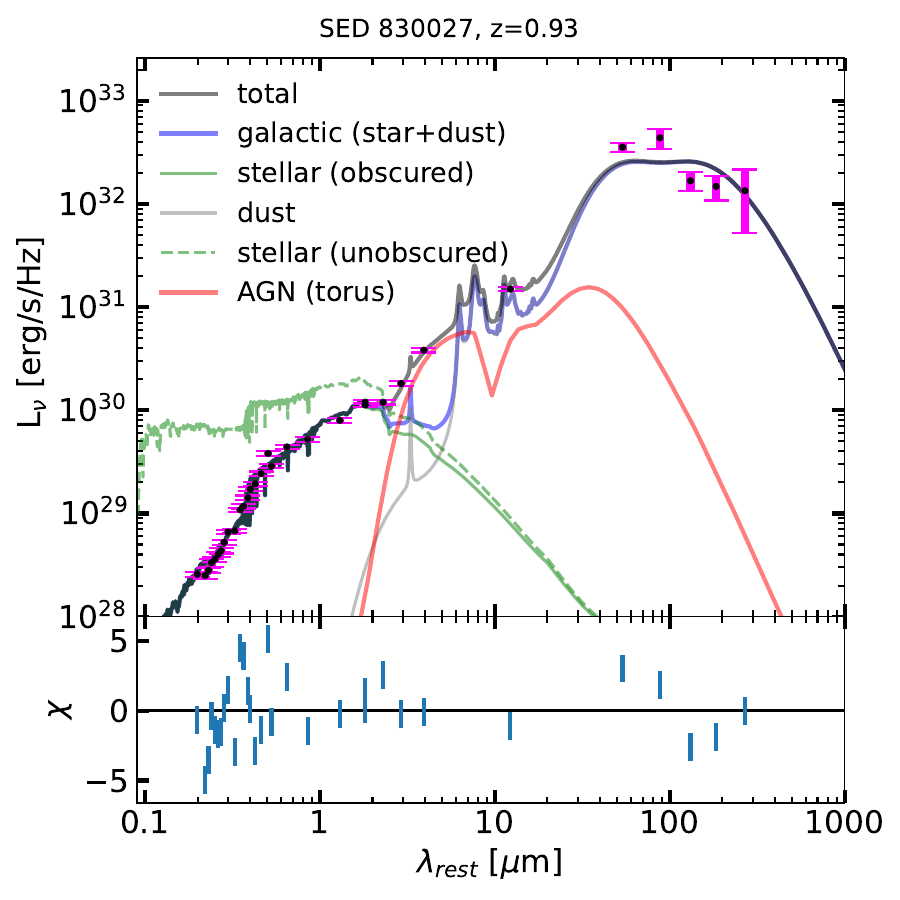}}
  \caption{Example SED fit for a \nev-selected AGN of the current sample. The black points with purple error bars are the photometric data, the dark gray line is the best-fit SED, composed of the galactic emission (blue line) and the AGN component (red line). The light grey line is the host galaxy dust emission, the continuous (dashed) green line refers to the dust-obscured (intrinsic) stellar emission. The fitting provides a stellar mass of $\log (M_{*}/\text{M}_{\odot})=10.64\pm0.05$, a SFR of $ 51\pm 10\,\rm{M_\odot\,yr^{-1}}$, and an AGN bolometric luminosity of $\log (L_{\text{bol}}^{\text{sed}}/\text{erg\,s}^{-1})=45.6_{-0.1}^{+0.3}$. The bottom panel shows the residuals, where $\chi=\,$($\,$observation$\,-\,$model$\,$)/error.}
  \label{fig:sed}
\end{figure}

Figure~\ref{fig:sed} shows the SED-fitting results for one of the \nev\ AGNs, source zCOSMOS 380027, undetected in the X-ray, at $z=0.9307$.
\subsubsection*{AGN bolometric luminosity}
Unlike type 1 objects, for obscured AGNs it is not possible to obtain the AGN bolometric luminosity from the optical emission. However, as the SED-fitting algorithm allows us to disentangle the AGN and the host contributions, one of the direct outputs of \textit{SED3FIT} is the AGN bolometric luminosity, derived via the integration of the spectrum of the central source that illuminates the torus.\par
We found that for 18 sources the SED fitting shows a low contribution of the AGN to the total emission, hence for these sources we have only upper limits on their AGN bolometric luminosity. The remaining 76 sources have a median $\log (L_{\text{bol}}^{\text{sed}}/\text{erg$\,$s$^{-1}$})=44.4 \pm 0.7$.\par
\begin{figure}
  \centering
  \resizebox{\hsize}{!}{\includegraphics{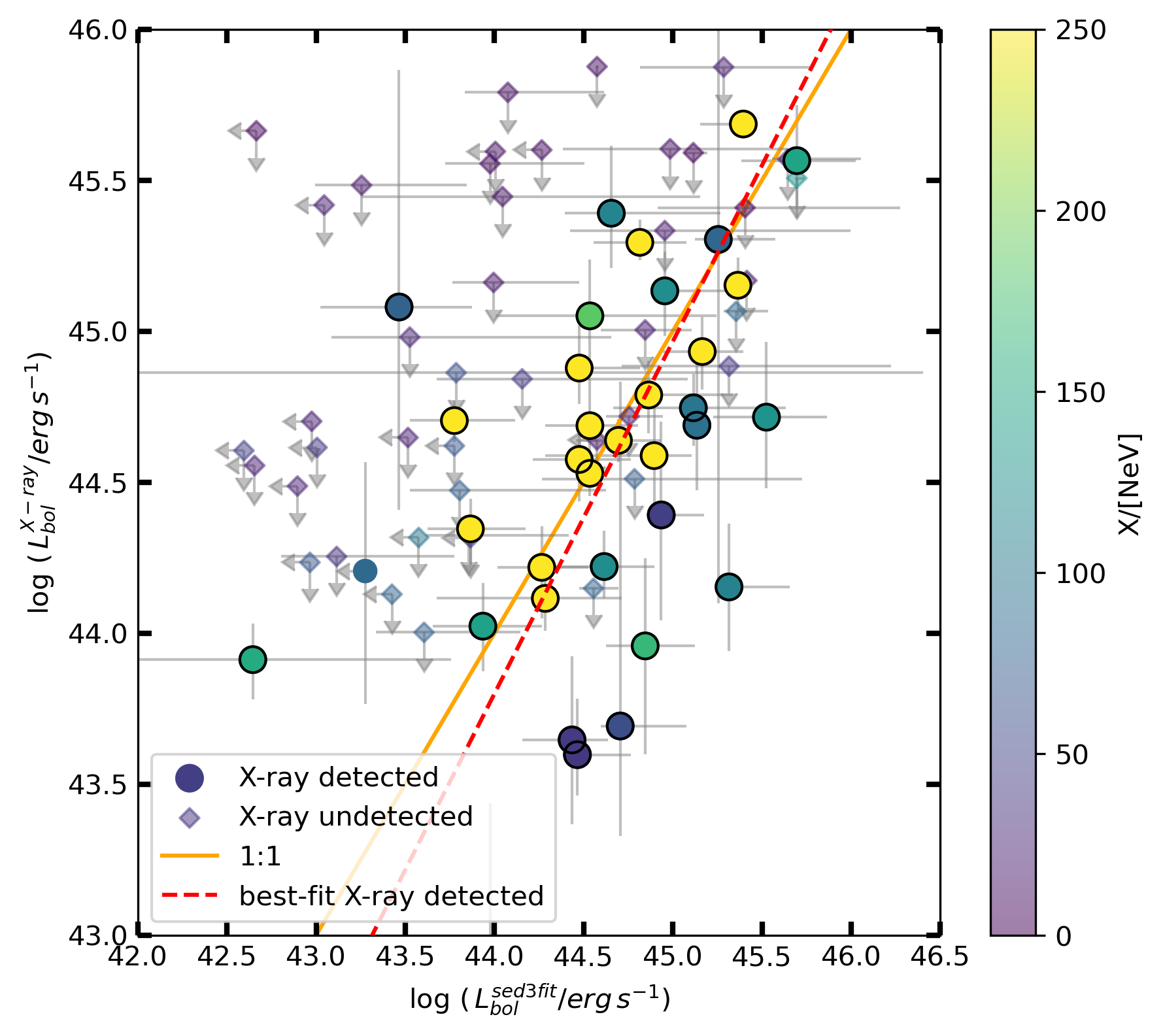}}
  \caption{Comparison of the AGN bolometric luminosities obtained from the SED-fitting (L$_{\text{bol}}^{\text{sed}}$) and from X-ray spectral analysis (L$_{\text{bol}}^{\text{x}}$). The L$_{\text{bol}}^{\text{x}}$ were computed from the $2 - 10$ keV rest-frame intrinsic luminosities using the \citet{lusso12} bolometric correction. Circles indicate X-ray-detected sources, while diamonds are the X-ray-undetected. For these sources, we have only an upper limit on their L$_{\text{bol}}^{\text{x}}$. The color code indicates the X/\nev\ ratio for the X-ray-detected sources and upper limits for the X-ray-undetected sources. The \textit{orange line} is the 1:1 correlation. The \textit{red line} is the best-fit relation for the X-ray-detected sources obtained via orthogonal distance regression.}
  \label{fig:lbol_comp}
\end{figure}

 To further confirm the validity of these luminosity estimates, we compared them with the bolometric luminosities obtained from the X-ray spectral analysis. We will refer to the bolometric luminosities computed from the SED-fitting as $L_{\text{bol}}^{\text{sed}}$ and to those obtained from X-ray analysis as $L_{\text{bol}}^{\text{x}}$. For the sources with X-ray detections, we were able to compute the intrinsic $2 - 10$ keV rest-frame luminosity, as reported in \S~\ref{cap_x_detec}. Using the bolometric correction $K_{\text{bol}}$ from \citet{lusso12} (see also \citealt{duras20}), we obtained the AGN bolometric luminosities L$_{\text{bol}}^{\text{x}}$ of these sources, with a median value of $ \log ( L_{\text{bol}}^{\text{x}} / \text{erg s}^{-1})=44.7 \pm 0.5$. Considering the X-ray-detected sources, the median (and 16-84$^\text{th}$ percentiles) L$_{\text{bol}}^{\text{sed}}/$L$_{\text{bol}}^{\text{x}}$ ratio is $1.00_{-0.01}^{+0.02}$. In Fig.~\ref{fig:lbol_comp} we show the comparison of the bolometric luminosities: except for two sources, the two bolometric luminosities are compatible within twice their uncertainties, with a Pearson correlation value $\rm{r}=0.46$ and a \textit{p}-value of $0.006$ (corresponding to a confidence $>2.7\,\sigma$). We performed an orthogonal distance regression and found a best-fit line (\textit{red line}) with slope $m=1.16 \pm 0.17$ and a dispersion of 0.6 dex. We note that \citet{stemo20} found a similar scatter of 0.6 dex between $L_{\text{bol}}^{\text{sed}}$ and $L_{\text{bol}}^{\text{x}}$ for its sample of $\sim2100$ AGNs across a wider range of luminosities. We note that the three detected sources on the bottom right of the plots are among the most obscured AGNs (lowest X/\nev\ ratios). For these sources, the lower X-ray-derived bolometric luminosities are likely linked to an underestimation of the X-ray luminosity due to the high obscuration. Further causes of the dispersion between the two bolometric luminosities could be linked to the assumed bolometric correction and to the fact that the X-ray emission can be affected by the AGN short-time variability (the X-rays tracing the innermost part of the AGN emission), absent in the SED-fitting derived bolometric luminosity (as the SED-fitting relies on the torus emission). In Fig. \ref{fig:lbol_comp}, we included the upper limit on the bolometric luminosities for the X-ray-undetected sources (\textit{diamonds}). In §\ref{sec:xray_undetected}, we discussed how we computed the upper limit on the $L_{\rm{bolo}}^{\rm{X-ray}}$ of these sources. We note that, similarly to the X-ray-detected sample, the X-ray-undetected AGNs with the highest X/\nev\ ratios (i.e. the less obscured ones) are the nearest to the 1:1 correlation, indicating that the obscuration should be the most prominent source of uncertainties in the $L_{\rm{bolo}}^{\rm{X-ray}}$. Considering only the X-ray-detected sample, we found that the bolometric luminosities derived from X-rays and SED-fitting are surprisingly similar, in particular, if we consider the low number of X-ray net counts and the degeneracies between the AGN and SF components in the SED-fitting. This correlation between the bolometric luminosities derived from the X-ray spectral analysis and the SED-fitting is extremely important as it confirms the reliability of both the SED-fitting procedure and the X-ray analysis allows us to use the estimate of the AGN bolometric luminosities derived from optical-to-FIR photometric data for the X-ray-undetected sources.\par
 
\begin{figure}
  \centering
  \resizebox{\hsize}{!}{\includegraphics{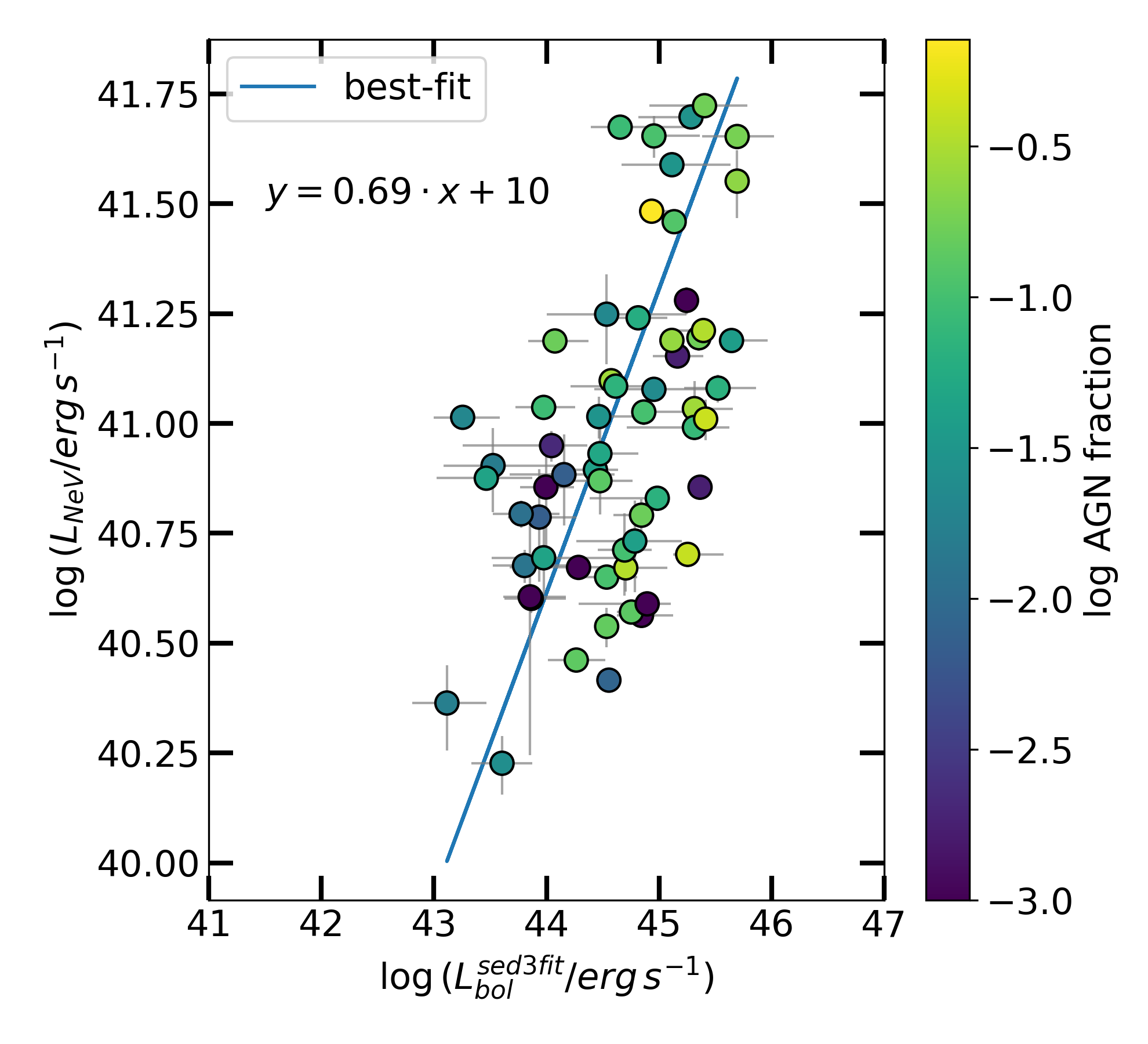}}
  \caption{ Luminosity of the \nev\ line as a function of the AGN bolometric luminosity obtained from the SED-fit. We excluded the 18 sources with only upper limits on their $L_{\rm{bol}}^{\rm{sed3fit}}$. The blue line is the best-fit relation obtained via an orthogonal distance regression. The \nev\ sources are colour coded on the basis of their AGN fraction measured in the $8-1000\,\mu\rm{m}$ wavelength range.}
  \label{fig:lnev_lbolo}
\end{figure}

As we can see from Fig.~\ref{fig:lnev_lbolo}, we also found a correlation (Pearson $\rm{r}\sim0.6$ with a $p\rm{-value}=7\times 10^{-6}$, i.e., significance at $\sim 4.5\,\sigma$) between the \nev\ line luminosity and the AGN bolometric luminosity. The \nev\ luminosity is aperture-corrected, but it is not corrected for the extinction in the host-galaxy, as it has been shown by M13 the mean optical extinction, derived from the observed \Hbeta/\Hgamma\ flux ratio, is low $(\langle E(B-V)\rangle = 0.18)$. This relation was expected due to the fact that the \nev\ in the NLR (where this line originates from) should be excited by the emission of the AGN central engine and thus, it should be a proxy of the AGN intrinsic power \citep[see also][]{gilli10,mignoli13,gruppioni16}. We performed an orthogonal distance regression (in the $\log-\log$ space) and found a best-fit linear  relation $\log{(L_{\rm{\nev}}/\rm{erg\,s^{-1}})}=0.69\times \log{(L_{\rm{bol}}^{sed3fit}/\rm{erg\,s^{-1}})} +10$. We excluded from this fit the 18 sources for which we had only an upper limit on their bolometric luminosity, caused by \texttt{sed3fit} choosing a best-fit model with a low AGN contribution to the total SED. The  $\L_{\rm{\nev}}-L_{\rm{bol}}$ relation, alongside the two similar relations found for a local sample by \citet{gruppioni16} using the IR \Nevir\ lines, confirm that the \nev\ emission is a good tracer of the AGN intrinsic emission and that it can be used to constrain the AGN bolometric luminosity both in the local universe and at $z\sim 1$.

\subsubsection*{AGN significance}
The SED-fitting procedure allowed us to separate the contribution of the AGN from that of the galaxy. However, this process is subject to a certain intrinsic degeneracy: an overestimation of the AGN fraction will result in an underestimation of the IR emission from the galaxy and thus of the SFR and vice versa. To further assess the reliability of the chosen torus models, as well as to estimate the importance of the AGN component on the total emission on a "per source" basis, we estimated the AGN significance using a Fisher test (F-test) between the best-fit $\chi^2$ with and without AGN component similarly to what was done in \citet{delvecchio14}, but see also \citet{bevington03}. We carried out a second run of \textit{sed3fit}, using the same optical and IR models, without any torus model. We compared the obtained $\chi^2_{\text{NO-AGN}}$ with the $\chi^2_{\text{AGN}}$ obtained using the torus models. The F-value was computed as
\begin{equation}
F_{\text{test}}=\frac{\chi ^2 _{\text{NO-AGN}}-\chi ^2 _{\text{AGN}}}{\bar \chi ^2 _{\text{AGN}}}
,\end{equation}
 where $ \bar \chi ^2 =\chi ^2/ dof$ and $dof$ is the number of degrees of freedom. As $dof$ we used the number of photometric points, in the case of the model without AGN, and the number of photometric points minus one, to take into account the additional parameter (the torus), for the model with AGN. Formally, the computed value of the F-statistic corresponds to the F-value of the ``ANOVA F-test in regression analysis of nested non-linear models" and the AGN significance is simply the \textit{p}-value probability converted to Gaussian standard deviation.\par
We obtained seventy-four sources (79\%) with an AGN significance $ \ge 1\,\sigma$,  forty-eight sources (51\%) with $\ge 2\,\sigma$, thirty-three (35\%) with $ \ge 3\,\sigma$ and twenty-four (26\%) with $ \ge 4\,\sigma$. Considering only the best value of the chi-squared, thirteen sources (14\%) were better fitted with a model without an AGN component; seven of them were sources for which we had only upper limits on their AGN IR luminosity, due to the low contribution of the AGN to the total SED. The fact that nearly half of the sample has an AGN significance $ < 2\,\sigma$ can be attributed to intrinsic low torus luminosities, with the torus emission largely diluted in the host-galaxy emission. In this regard, all thirteen sources that were better fitted without the AGN component were not detected in the X-ray. Moreover, all the objects with AGN significance $ \le 1\,\sigma$ are not X-ray-detected, except for three sources. Two of these, however, have luminosity in the lower end of our distribution (L$_{2-10\text{keV,intr}}<10^{43}$ erg/s).\par
 We studied the distributions of the \nev\ luminosities for the sources with AGN significance $ < 2\,\sigma$  and those with AGN significance $ \ge 2\,\sigma$. We found a segregation of the sources with low AGN significance at low \nev\ luminosities. In fact, while $44\%$ of the sources with $ \ge 2\,\sigma$ AGN significance has low \nev\ luminosity ($L_{\text{\nev}}<10^{41}$ erg/s), this percentage goes up to $85\%$ for the sources with AGN significance $ < 2\,\sigma$. This segregation is in support of the fact that low AGN luminosities (the \nev\ emission is a proxy of the nuclear intrinsic emission) may be challenging in separating the AGN component from the galaxy emission.

\subsubsection*{Stellar mass}
The median (and 16-84$^\text{th}$ percentiles) stellar mass is $\log{(\text{M}_*\,/\,M_{\odot})}=10.91_{-0.46}^{+0.28} $. We noted that the stellar masses are extremely well constrained by the SED-fitting procedure with a mean $\Delta\log\text{M}_{*}\approx 0.06$. We did not find any significant trend of the stellar mass with the X-ray detection.

\subsubsection*{Star formation rate}
\begin{figure}
  \centering
  \resizebox{\hsize}{!}{\includegraphics{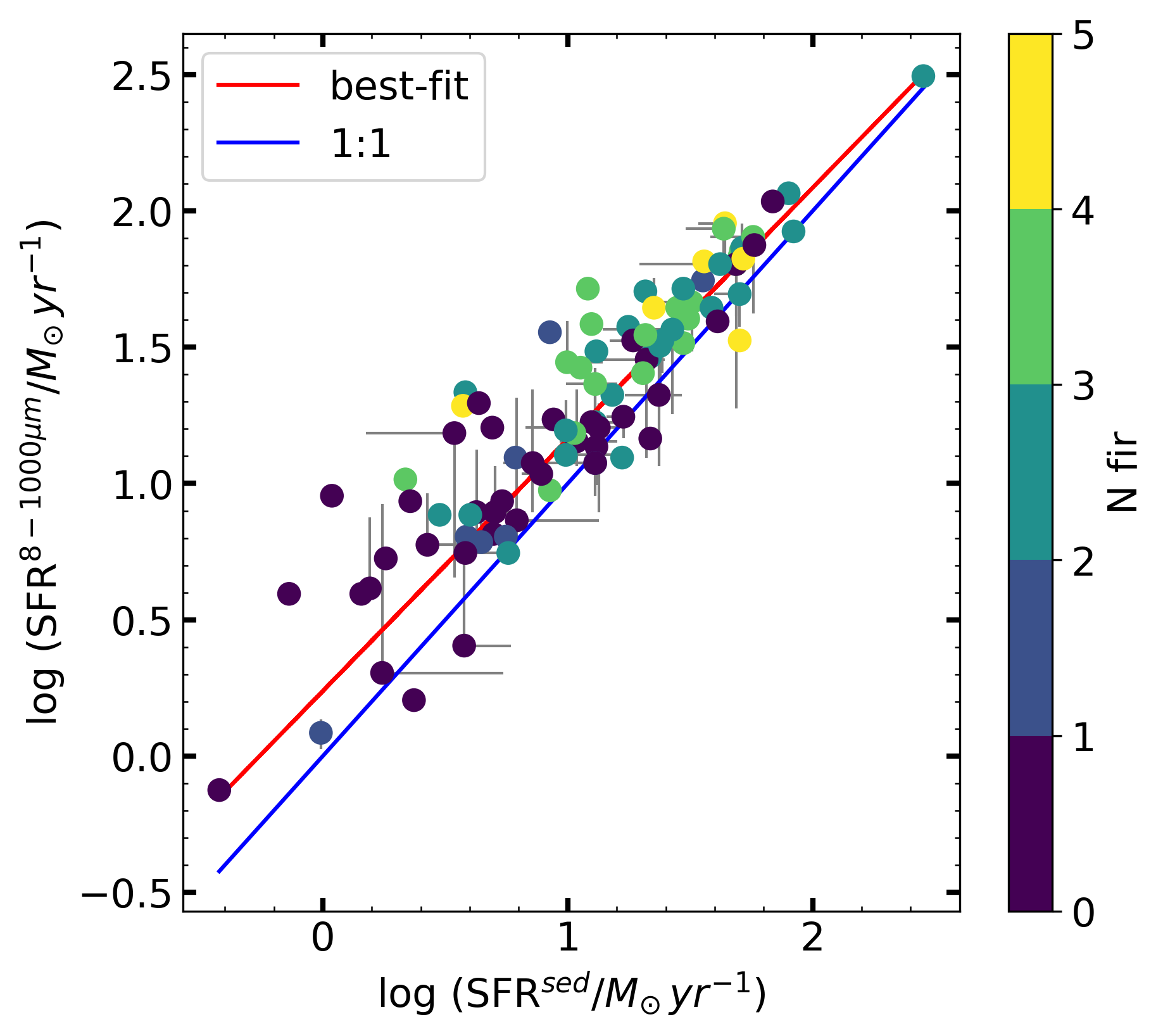}}
  \caption{Comparison between the SFR obtained from the optical-NIR bands with those from FIR. SFR$^{\text{sed}}$ are derived through the modeling of the stellar emission in the UV-to-NIR. The color code indicates the number of photometric detections for each source in the FIR band. A low number of FIR detections may influence the goodness of the FIR SED-fitting, hence the SFR$^{8-1000\mu\text{m}}$, but we did not find any trend between the SFR$^{8-1000\mu\text{m}}-$SFR$^{\text{sed}}$ offset and the number of IR photometric detections. The blue line is a 1:1 line; the red line is the best-fit line with a slope of $m=0.92$ and $c=0.24 $, in the $\log (\textnormal{SFR}^{8-1000\mu\text{m}}/M_\odot \, yr^{-1})=c+m\,\log (\text{SFR}^{\text{sed}}/M_\odot \, yr^{-1})$ notation.}
  \label{fig:sfr}
\end{figure}

The SFR of our sample was obtained in two different ways: using the best-fit model (SFR$^{\text{sed}}$) and using the $8 - 1000\,\mu$m SF luminosity (SFR$^{8-1000\mu\text{m}}$).
The SFR$^{\text{sed}}$ is the mean SFR of the last 0.01 - 0.1 Gyr as obtained from the modeling of the stellar component in the UV-to-NIR regime with \textit{sed3fit}. The code uses the UV-optical-NIR library of \citet{bruzual07}, which produced the optical-to-NIR spectra, by considering the spectral evolution of stellar populations for different metallicities and star formation histories and assuming a Chabrier IMF \citep{chabrier03}. We obtained a median SFR$^{\text{sed}}= 12.9_{-9.1}^{+30.4}\,M_{\odot}\,yr^{-1}$ ($\Delta\text{SFR}/\text{SFR}\approx 0.6\%$). The SFR$^{8-1000\mu\text{m}}$ is the SFR averaged over the last 100 Myr computed by the emission of dust heated by young stars as well as of evolved stellar populations. It is derived from the IR luminosity, once the AGN contribution is removed, exploiting the \citet{kennicutt98} relation (which assumes a Salpeter IMF) $\textnormal{SFR}^{8-1000\mu\textnormal{m}}\, (\textnormal{M}_{\odot}/\textnormal{yr})\,=\,4.5\,\times 10^{-44}\, \textnormal{L}_{8-1000\mu\textnormal{m}}\, (\textnormal{erg/s})$ and using $ \textnormal{SFR}_{\textnormal{Chabrier}}\,=\,0.67\,\times\textnormal{SFR}_{\textnormal{Salpeter}} $ to convert it to a Chabrier IMF. The median SFR is SFR$^{8-1000\mu\text{m}}= 20.4_{-9.2}^{+43.4}\,M_{\odot}\,yr^{-1}$. Because the SFR$^{8-1000\mu\text{m}}$ is heavily dependent on the fitting of the FIR band, the reliability of this value is linked to the accuracy at which the far-IR is measured. As we can see from Fig.~\ref{fig:sfr}, in which we compared the two SFR values for each source, we have a systematic offset of $\sim 0.15\,$dex, but we did not find a clear correlation between this offset and the number of IR photometric detections nor with the SFR uncertainties.   

\subsubsection*{SFR-M$_{*}$ relation}
\begin{figure}
  \centering
  \resizebox{\hsize}{!}{\includegraphics{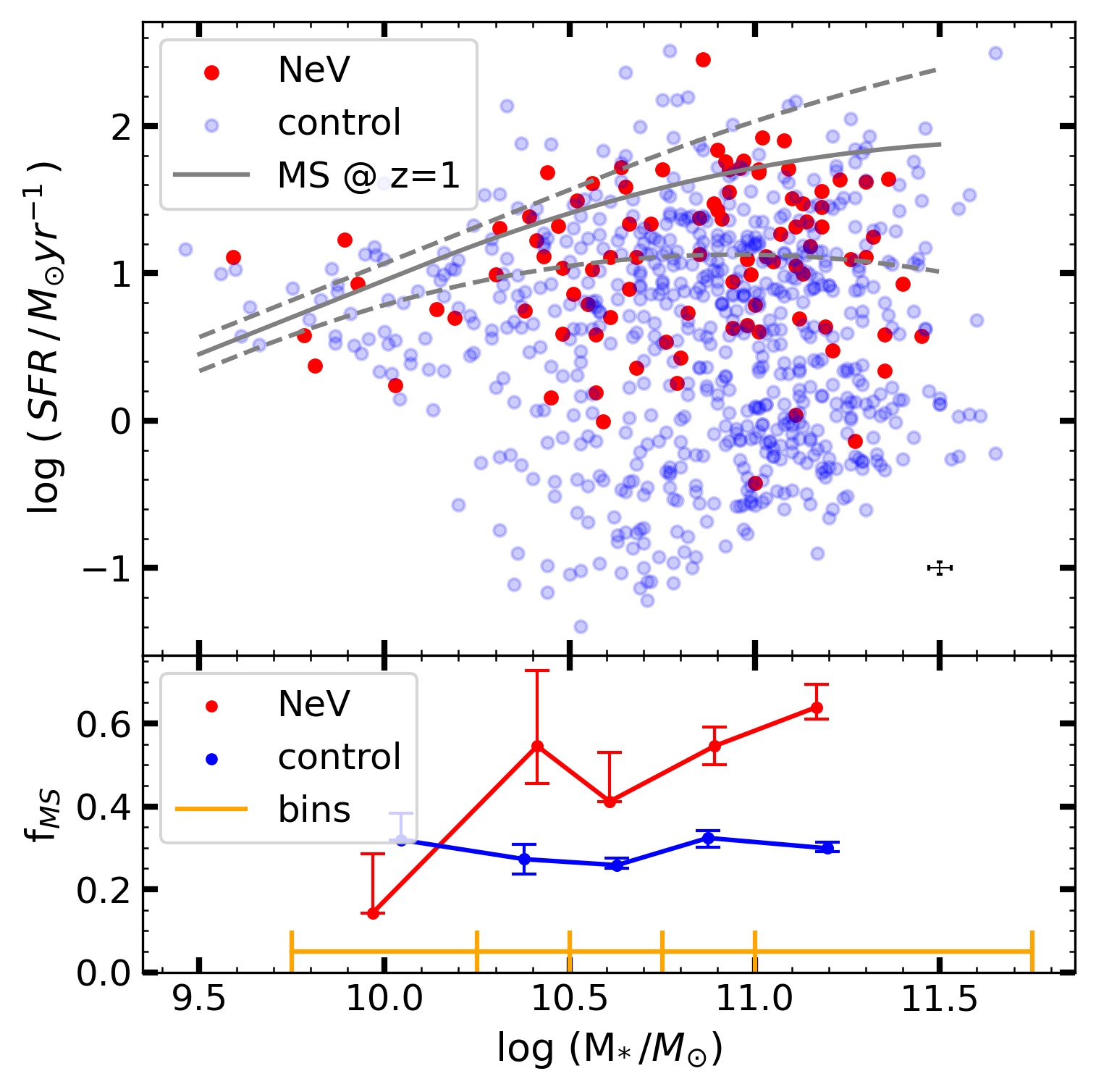}}
  \caption{Comparison of the positions of the \nev{} (\textit{red}) and stellar mass- and redshift-matched control (\textit{blue}) samples in the SFR$-$M$_*$ plane. \textit{Upper panel}: Grey solid line is the \citet{schreiber15} MS, the grey dashed lines its $1\,\sigma$ dispersion. The black error bars in the bottom right are the mean uncertainties for the \nev{} sample sources. \textit{Bottom panel}: Fraction of sources within the $1\,\sigma$ dispersion of the MS. In orange: the adopted stellar mass binning.}
  \label{fig:sfr_ms_parent}
\end{figure}

We investigated whether the hosts of the \nev{}-selected AGNs lie within the SFR-M$_{*}$ main sequence \citep{noeske07}, as shown in  Fig~\ref{fig:sfr_ms_parent} (\textit{red dots}). We used the \citet{schreiber15} main sequence (MS), in which the SFR is a function of both the stellar mass and the redshift:
\begin{equation} 
\log{(\text{SFR}_{\text{MS}}/\text{M}_{\odot}\,\text{yr}^{-1}])}=\,m-m_0+a_0r-a_1[\max (0,m-m_1-a_2r)]^2
\label{eq:MS}
,\end{equation}
where $r\equiv \log(1+z)$, $m \equiv \log (\text{M}_{*}/10^9 \text{M}_{\odot})$, $m_0=0.5 \pm 0.07$, $a_0=1.5 \pm 0.15$, $a_1=0.3 \pm 0.08$, $m_1=0.36\pm 0.3,$ and $a_2=2.5 \pm 0.6$.
Using the masses obtained from the SED-fitting and the \citet{schreiber15} SFR-M$_{*}$ relation (eq \ref{eq:MS}), we obtained a median SFR$_{\text{MS}}= (\, 33.2_{-14.8}^{+21.0}\,)\, \textnormal{M}_{\odot}/\textnormal{yr}$. We used the SFR$_{\text{norm}}=$SFR$^{\text{sed}}/$SFR$_{\text{MS}}$ to trace how much a source deviates from the MS. We found a median SFR$_{\text{norm}} = 0.48_{-0.37}^{+0.76}$, which indicates that a significant fraction our sample has a SFR lower than what is expected for SF galaxies.\par
We investigated if the SFR and the position in the SFR-M$_{*}$ plane depend on the number of IR photometric detections (which are fundamental in constraining the overall SED in \textit{sed3fit}). For the sources with at least one IR detection, we performed another run of SED-fitting without the IR photometric points. The comparison of the SED-derived properties with those obtained using all the photometric points shows larger uncertainties in the parameters for the fits without the IR, but no systematic effect is present.\par
M13 performed a morphological classification of the \nev{} sample and a comparison with the morphologies of a control sample. The galaxies were classified following \citet{nair2010} using the F814W-band images. The position of the \nev{} in our SFR-M$_*$ plane, with the majority of the galaxies within the MS, is in agreement with the M13 morphological classification, in which early-spirals are the most commonly found type for the \nev{} sample (rather than irregular or late-spirals). Moreover, M13 found a lower fraction of elliptical galaxies with respect to their control sample; this is in agreement with the paucity of low-sSFR galaxies we found in the \nev{} sample (see \S~\ref{sec:parent}).

\subsubsection*{SFR$-$L$_{\rm{bolo}}$ relation}
We investigated the relation between host galaxies and AGNs by comparing the bolometric luminosity of the AGN with the SFR and the SF-related IR luminosity. We found a statistically significant but mild correlation between the SFR and the $L_{\rm{bolo}}$ (Pearson $\rm{R}=0.28$ and $P$-value $=0.02$, corresponding to $2.3\,\sigma$), and a stronger correlation between $L_{\rm{IR,SF}}$ and $L_{\rm{bolo}}$ ($\rm{R}=0.44$ and a $P$-value $=1.6\times 10^{-4}$).\par
\begin{figure}
  \centering
  \resizebox{\hsize}{!}{\includegraphics{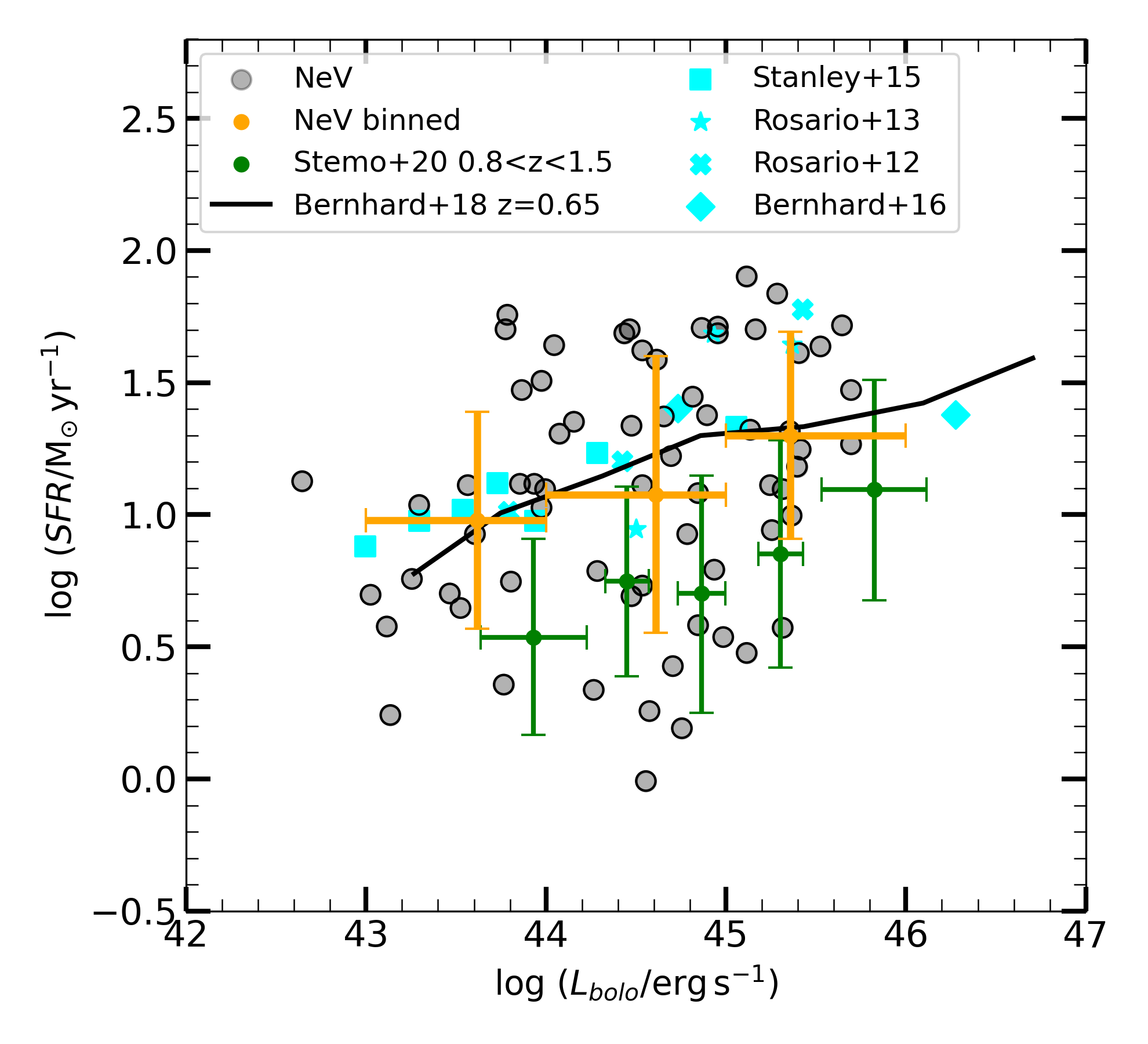}}
  \caption{SFR as function of the AGN bolometric luminosity for the \nev-selected AGNs (\textit{red circles}). The average SFR in three bins of SFR is reported in \textit{orange}, with the error bar representing the $16^{\rm{th}}$ and $84^{\rm{th}}$ percentiles. The \textit{green points} are the average SFR for the sample of X-ray and IR -selected AGNs of \citet{stemo20}, while the \textit{cyan symbols} represent those from \citet{rosario12,rosario13,stanley15,bernhard16} collected by \citet{bernhard18}. The \textit{black line} is the prediction from their  mass-dependent model.} 
  \label{fig:lbolo_sfr}
\end{figure}

In Fig.~\ref{fig:lbolo_sfr}, we show the position of the \nev\ AGNs (\textit{red dots}) in the $L_{\rm{bolo}}-$SFR plane and the comparison with the AGN sample of \citet{stemo20} and with the model prediction of \citet{bernhard18}. The orange points correspond to the median (and $16^{th}$, $84^{th}$ percentiles) SFR of the \nev\ sample binned in AGN bolometric luminosity. The green points are the average SFR binned in $L_{\rm{bolo}}$ for a sample of 2585 X-ray and IR selected AGNs in the $0.2 < z < 2.5$ regime from \citet{stemo20}. In the $0.8-1.5$ redshift range, we found a similar flat evolution of the SFR with the AGN bolometric luminosity, however with $\sim 0.4$ dex higher SFR. We speculate that this could be due to the different selection techniques, as \citet{stemo20} found that most of their sample is composed of quiescent galaxies, while in ours they are $<10\,\%$. These differences could be related to the band used for the selection (i.e., X-ray or IR), as the authors note that it impacts their SFR.\par
We also plot the published relationships between X-ray luminosity and SFR of \citet[][\textit{cyan symbols}]{rosario12,rosario13,stanley15,bernhard16} collected by \citet{bernhard18}, alongside the prediction of their mass-dependent model (\textit{black lines}). We used the \citet{lusso12} bolometric correction to compute the $L_{\rm{bolo}}$ from the AGN X-ray luminosity, and 'converted' the SFR to a \citet{chabrier03} IMF. As we can see from Fig.~\ref{fig:lbolo_sfr}, our results are extremely similar to the observed relationships and model predictions.

\subsubsection*{Eddington ratio}
An illustrative way of estimating the SMBH growth is via the Eddington ratio $\eddrat=\text{L}_{\text{bol}}/\text{L}_{\text{Edd}}$: the ratio between the AGN luminosity and the theoretical maximum luminosity that can be emitted by the AGN (determined by the balance between radiation pressure and the BH gravitational attraction) and is therefore an indicator of the small scale accretion process. According to the in situ co-evolution model (see \S~\ref{sec:insitu}) the Eddington ratio is strictly connected to the phase of the AGN: the highly-accreting obscured AGNs should have  $\eddrat \gtrsim 1$, while once the AGN feedback kicks in and starts removing the gas from the host-galaxy, a transition to sub-Eddington regimes ($\eddrat \lesssim 0.1$) should occur. AGNs may accrete in multiple short burst phases \citep[e.g.,][]{Schawinski15}, in this case, the AGN would ``quickly'' ($\tau\sim 10^5\,\rm{yr}$) pass through high- and low- Eddington ratios multiple times.\par
\begin{figure}
  \centering
  \resizebox{\hsize}{!}{\includegraphics{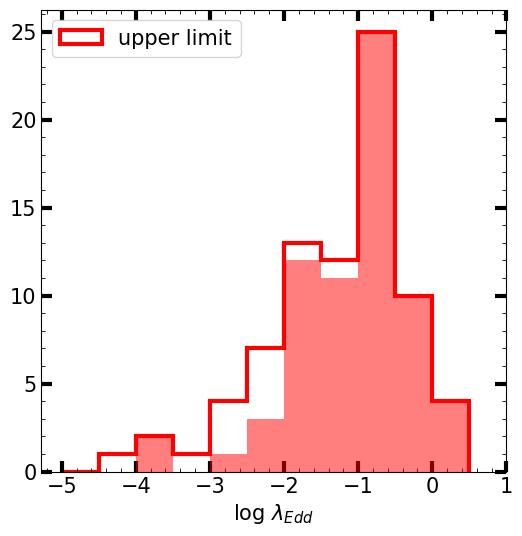}}
  \caption{Eddington ratio distribution of the \nev-selected sample of type 2 AGNs. The $\eddrat$ was computed with the M$_{BH}$ obtained from the stellar mass (using the \citealt{suh20} M$_{\text{BH}} - \text{M}_{*}$ relation) and with the AGN bolometric luminosity from the SED fitting. The sources for which we had only an upper limit on their bolometric luminosity are reported as a white histogram. $40\%$ of the sources have $\eddrat \ge 0.1$, suggesting that a significant fraction of the \nev{} sample are in a highly accreting phase.} 
  \label{fig:eddratio}
\end{figure}

We used the \citet{suh20} $M_{\text{BH}} - \text{M}_{*}$ relation to estimate the SMBH masses using the stellar masses obtained from the SED-fitting. This relation was obtained from a sample of 100 X-ray selected AGNs in the COSMOS field with host galaxy masses from SED-fitting decomposition and BH masses computed considering single epoch H$\alpha$, H$\beta$, and \mgii\ broad line widths and line/continuum luminosity as a proxy for the size and velocity of the broad-line region (BLR). We obtained a median $\log{(\text{M}_{BH}/M_{\odot})}=7.5_{-0.7}^{+0.4}$. We compared the \mbh with those obtained using the \citet{reines15} and the \citet{shankar16} relations, finding values not significantly different (KS probabilities of 0.02 and 0.08, respectively). A more detailed comparison can be found in Appendix \ref{sec:mbhrelations}. 
From the \mbh, we computed the Eddington luminosity and used it with the AGN bolometric luminosity to estimate the Eddington ratio (Fig.~\ref{fig:eddratio}). We found a median $\lambda_{Edd}=0.12_{-0.10}^{+0.31}$. For 25 sources ($\sim 25\%$) we obtained an upper limit of their \eddrat only. Forty percent of the sources have $\eddrat \ge 0.1$, and $5\%$ are accreting near or above the Eddington limit. We must caution that the \eddrat is heavily dependent on the \mbh, and those were computed assuming the \citet{suh20} relation. Unfortunately, our current knowledge of $\mbh-\text{M}_{*}$ relation is mostly local and usually derived from unobscured AGNs (due to the fact that the \mbh estimates are obtained from the profile of broad emission lines). Moreover, the advent of JWST has recently highlighted the presence of several high-$z$ galaxies with overmassive BH with respect to local relations \citep[e.g.,][]{ubler23,harikane23,pacucci24} and opened the possibility of a redshift evolution of the M$_{\text{BH}} - \text{M}_{*}$ relations. In any case, using any of the three local investigated functions suggests that a significant fraction of the \nev{} sample are AGNs in a high accretion phase.

\subsubsection*{Molecular gas fraction}\label{sec:molgas}
We used the \citet{kaasinen19} relation (eq \ref{eq:kaasinen}) to estimate the molecular gas mass of the sample. They obtained this relation linking the L$_{850\mu m}$ to the CO luminosity, which, in turn, is a proxy of the molecular gas mass:  
\begin{equation}
\label{eq:kaasinen}
\text{M}_{molgas}^{850\mu m}\,\, (M_{\odot})= \left( \frac{\text{L}_{850\mu m}}{erg\,s^{-1}\,Hz^{-1}}\right) \left( \frac{1}{6.2 \times 10^{-19} \left( \text{L}_{850\mu m} / 10^{31}  \right)^{0.07}} \right) 
.\end{equation}
We computed L$_{850\mu m}$ from the host-galaxy component, derived via the SED-fitting. The median molecular gas mass of the sample was $\log (M_{molgas}^{850\mu m}/ M_{\odot})= 10.4 _{-0.5}^{+0.4}$. We advise taking these results with a bit of caution as they depend on the L$_{850\mu m}$ measured from the best-fit SED, for which the nearest filters at our disposal were the SCUBA/JCMT at $\lambda_{\rm{rest}}\sim 225\,\mu\rm{m}$ and the SPIRE/Herschel $\lambda_{\rm{rest}}\sim 250\,\mu\rm{m}$. However, we also found a very good agreement (with the exception of one source) between the dust mass measured from the  L$_{850\mu m}$ \citep[using a gas-to-dust mass ratio in the $100-200$ range, e.g.,][]{tacconi20} and those obtained as a direct output from the \textit{sed3fit} fitting, which relies on the fitting of the entire IR SED.\par
We used the $\text{M}_{molgas}^{850\mu m}$ to obtain the molecular gas fraction of the galaxies, defined as $f_{\text{mol}}=\text{M}_{molgas}^{850\mu m}/(\text{M}_{molgas}+\text{M}_*)$. The sample had a median $f_{\text{mol}}=0.24 _{-0.14}^{+0.30}$. We found that our sample has f$_{\text{mol}}$ in agreement with those in \citet{dessauges20}, who collected from the literature the f$_{\text{mol}}$ for CO detected main sequence galaxies at $0 \le z\le 6$. The $f_{\text{mol}}$ distribution indicates that the majority of the \nev\ sources have cold gas available to fuel the SF and, in contrast to the ``read and dead" elliptical galaxies, they will likely continue to forming stars, not having depleted their reservoir yet.

\section{Comparison with the control sample}\label{sec:parent}
\begin{figure}
  \centering
  \resizebox{\hsize}{!}{\includegraphics{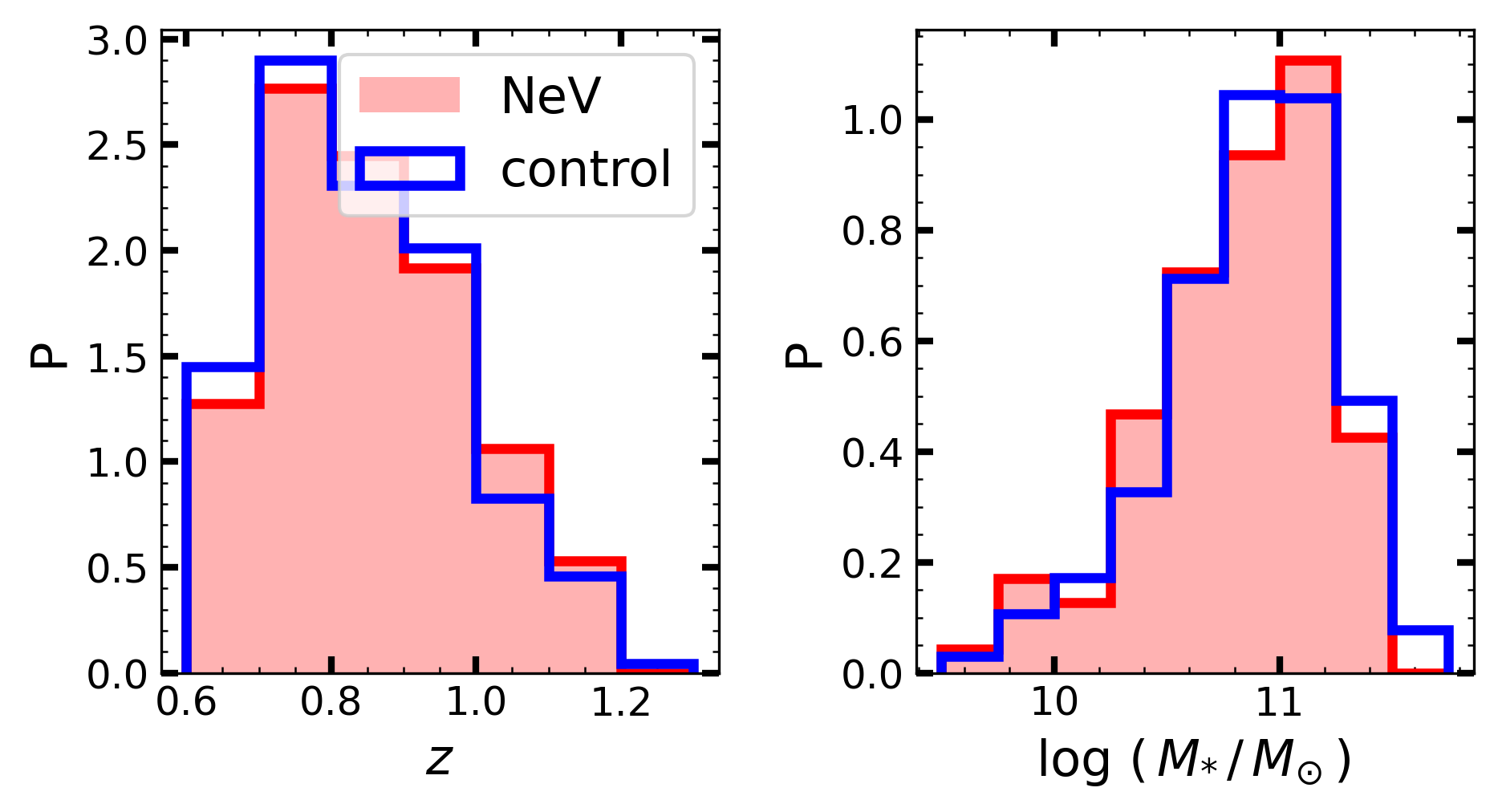}}
  \caption{Comparison of redshift (\textit{left}) and stellar mass (\textit{right}) distributions of the \nev{} (\textit{red}) and control (\textit{blue}) samples.}
  \label{fig:z_mstar_comparison}
\end{figure}

To be able to make a comparison between our sources and non-active galaxies in the COSMOS field, we built a control sample. For each source in the \nev{} sample, we selected eight sources in the zCOSMOS20k catalog, matched in redshift (with a $0.01-0.05$ accuracy) and stellar mass ($0.1\,$dex accuracy). From this sample, we excluded the 51 sources with the X-ray source flag in the COSMOS2020 catalog. We performed the SED-fitting in the same way as we did for the \nev{} sample (see Section \ref{sec:sed_fitting}) and excluded 30 sources for which the SED-fitting failed to properly reproduce the emission. We did not replace these sources as they were uniformly distributed both in stellar mass and redshift, and we still had a statistically adequate number of sources. The final control sample was composed of 618 sources, Fig. \ref{fig:z_mstar_comparison} shows the comparison of the $z$ and $M_*$ distributions between the \nev{}  and the control samples. In Table \ref{tab:comparison}, we report the comparison between the main properties of the \nev{} sample and of the control sample. \par

\begin{table}
\caption{Comparison of the main properties of the \nev{} sample and of the control sample. We report the median values (with the $16^{th}$ and $84^{th}$ percentiles) for the redshift $z$, stellar mass, SFR, specific SFR (sSFR=SFR/M$_*$), SFR$_{\text{norm}}$=SFR$/$SFR$_{\text{MS}}$, fraction of sources within the \citet{schreiber15} MS (f$_{\text{MS}}$), molecular gas mass (M$_{molgas}$),  the molecular gas fraction ( f$_{\text{mol}}$=M$_{molgas}/($M$_{molgas}+$M$_*)$), age of the galaxy (t$_{\text{age}}$, from the SED-fitting), BH mass (M$_{\text{BH}}$), and Eddington ratio ($\lambda_{Edd}$).}
\label{tab:comparison}
\centering
\begin{tabular}{ccc}
\hline \hline
 & \nev{} & Control    \\
\hline 
$z$ & $0.86_{-0.15}^{+0.11}$ & $0.85 \pm 0.11$ \\[3pt]
$\log{(\text{M}_*\,/\,M_{\odot})}$ & $10.91_{-0.46}^{+0.28} $ &$10.91_{-0.42}^{+0.30}$ \\[3pt]
SFR$^{\text{sed}}$ ($M_{\odot}\,yr^{-1}$) & $12.9_{-9.1}^{+30.4}$ & $4.3_{-3.7}^{+14.4}$\\[3pt]
$\log{(\text{sSFR}/yr^{-1})}$ & $-9.7 _{-0.7}^{+0.5} $ & $-10.2_{-1.1}^{+0.8}$ \\[3pt]
SFR$_{\text{norm}}$ & $0.48_{-0.37}^{+0.76}$ & $0.16_{-0.14}^{+0.56}$ \\[3pt]
f$_{\text{MS}}$ & $0.52_{-0.02}^{+0.08}$ & $0.30\pm 0.01$ \\[3pt]
$\log{(\text{M}_{molgas}\,/\,M_{\odot})}$  & $10.4 _{-0.5}^{+0.4}$ & $10.3_{-0.7}^{+0.5}$\\[3pt]
f$_{\text{mol}}$ & $0.24 _{-0.14}^{+0.30}$ & $0.20_{-0.15}^{+0.26}$ \\[3pt]
$\log{(\text{t}_{\text{age}}/yr^{-1})}$ & $9.48 _{-0.25}^{+0.20}$ & $9.63_{-0.30}^{+0.15}$ \\[3pt]
$\log{(\text{M}_{\text{BH}}/M_{\odot})}$ &  $7.5_{-0.7}^{+0.4}$&  \\[3pt]
$\lambda_{Edd}$ &  $0.12_{-0.10}^{+0.31}$&  \\[3pt]
\hline
\end{tabular}
\end{table}

\begin{figure}
  \centering
  \resizebox{\hsize}{!}{\includegraphics{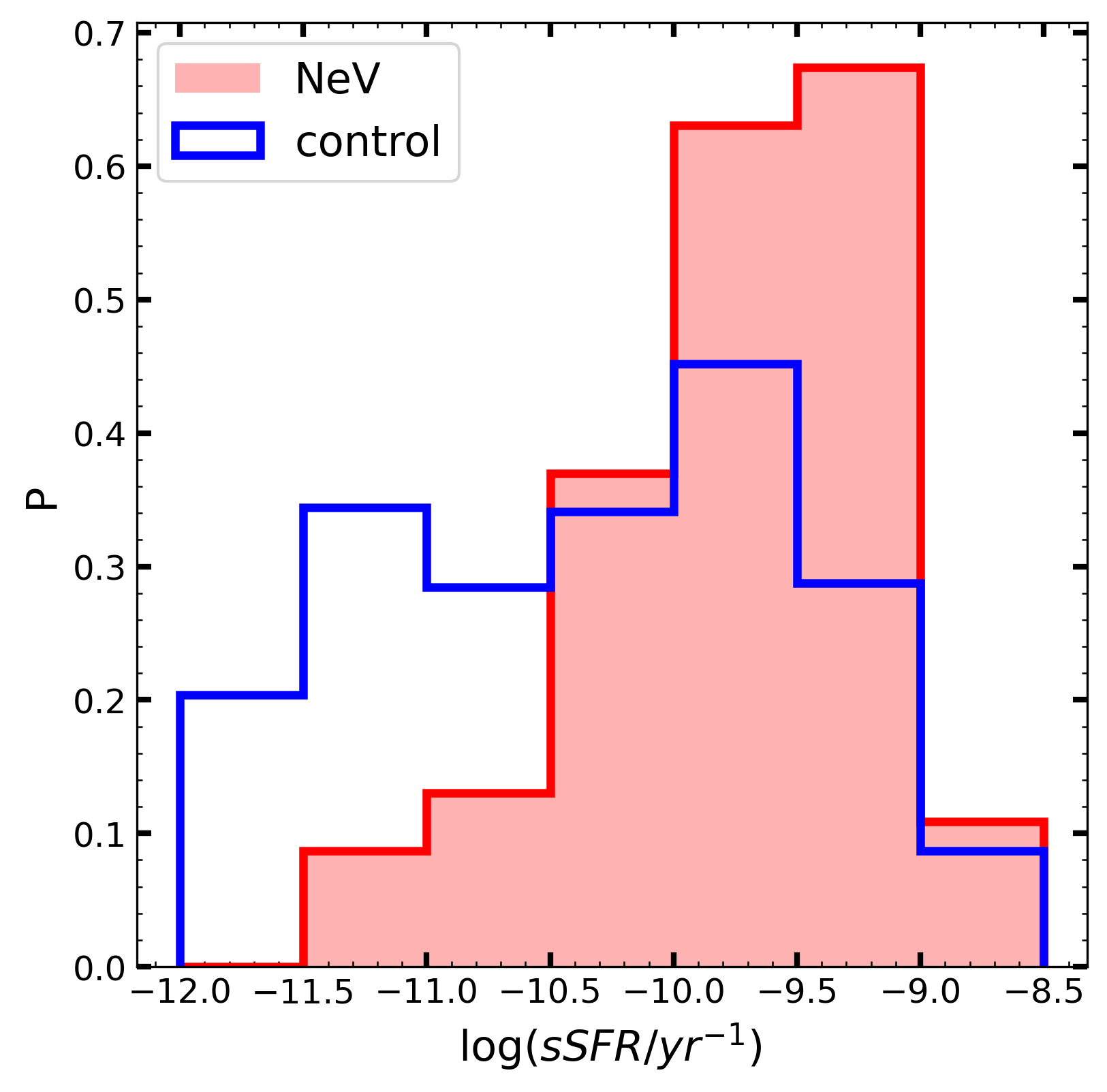}}
  \caption{Comparison of the specific SFR (sSFR=SFR/M$_*$) distributions of the \nev{} (\textit{red}) and control (\textit{blue}) samples. The two sSFR distributions are significantly different (KS-test P$=1.3\times 10 ^{-7}$), with the \nev{} sample lacking a population of ``quiescent'' galaxies ($\log(\textnormal{sSFR}/yr^{-1}) \leq -11$).}
  \label{fig:ssfr_parent}
\end{figure}

The \nev{} sample has, on average, a higher SFR with respect to the control sample, although both reach values as high as $\sim \, 300 M_\odot /yr$. Considering the position in the SFR-M$_{*}$ plane, we found that the \nev\ sample has a higher fraction of sources in the MS (i.e., a lower fraction of sources below). As we can see from Fig. \ref{fig:sfr_ms_parent}, for $\log{(\text{M}_*\,/\,M_{\odot})} \geq 10.25$, there is a clear difference in the fraction of sources within the MS, with the control sample having a constant fraction f$_{\textnormal{MS}}\sim 40\%$, while the \nev{} f$_{\textnormal{MS}}\sim 60\%$. This difference between the two samples appears more evident in Fig. \ref{fig:ssfr_parent}: the control sample specific SFR (sSFR=SFR/M$_*$) has a bimodal distribution, while the \nev{} has a higher fraction of sources at high sSFR and seems to lack a population of ``quiescent'' galaxies ($\log(\textnormal{sSFR}/yr^{-1}) \leq -11$). The two sSFR distributions are significantly different (KS-test P$=1.3\times 10 ^{-7}$). The comparison of the t$_{\text{age}}$ (the age of the oldest stars in the galaxy, as provided by \textit{sed3fit}) distributions shows a significant difference (KS-test P$\sim 6\times 10 ^{-7}$) between the two samples, with the \nev{} sources being, on average, younger. This is in agreement with the results from M13, who compared the morphologies of our sample with those of a mass-matched control sample of non-active galaxies and found that the \nev\ selection prefers early-type galaxies, with a lower fraction of late spiral and elliptical morphologies. \par
\begin{figure}
  \centering
  \resizebox{\hsize}{!}{\includegraphics{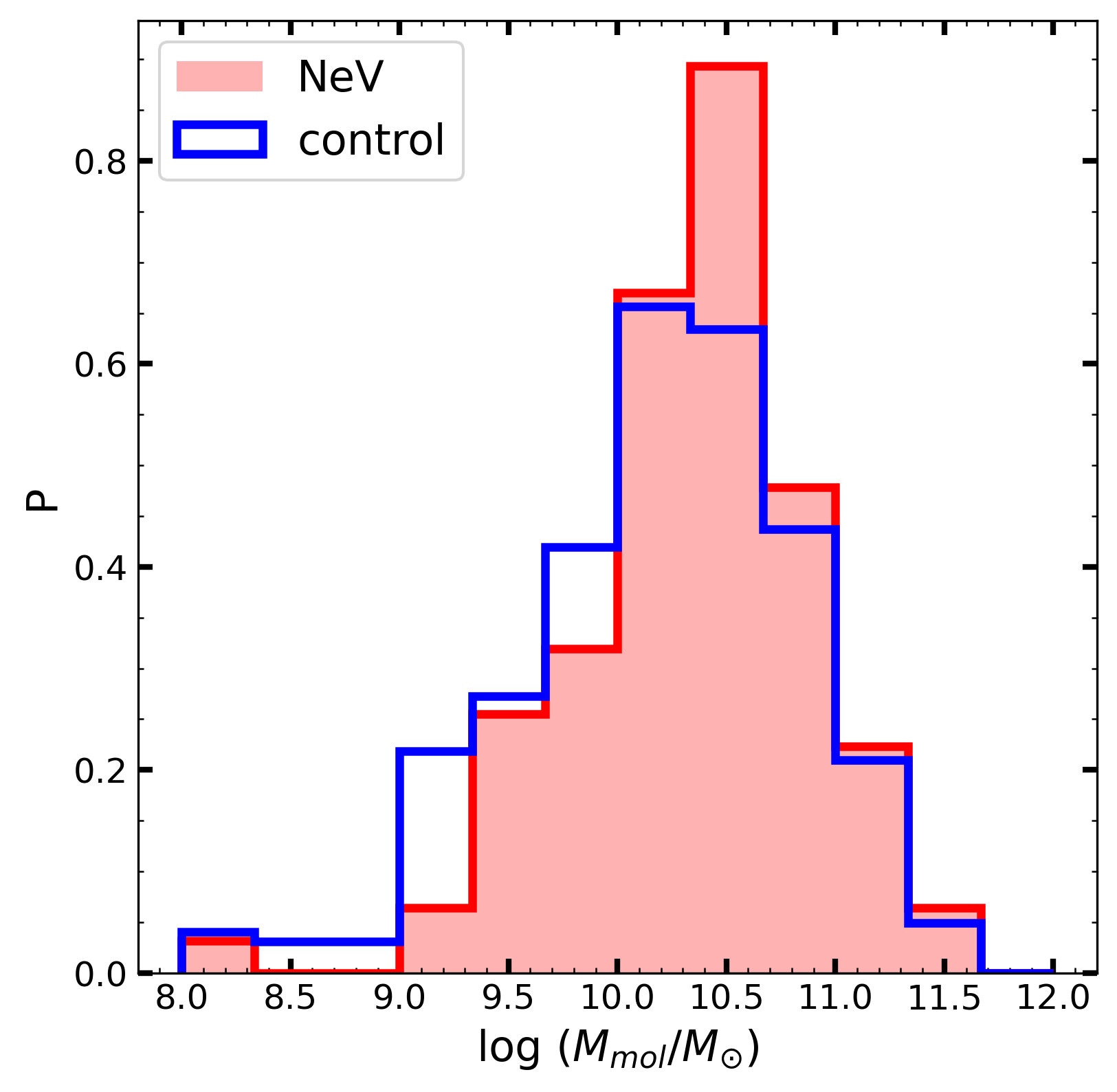}}
  \caption{Comparison of the molecular gas mass distributions of the \nev{} (\textit{red}) and control (\textit{blue}) samples. We obtained the M$_{\text{mol}}$ from the L$_{850\mu m}$, using the \citet{kaasinen19} relation. The two M$_{\text{mol}}$ distributions are not significantly different.}
  \label{fig:mmol_hist}
\end{figure}

We investigated the cold gas content of the galaxies to test whether the difference in sSFR (and the lack of ``quiescent'' galaxies) originated from a different amount of gas available to fuel the SF. As in \S~\ref{sec:molgas}, we computed M$_{\text{molgas}}$ and f$_{\text{mol}}$ for both samples. We did not find any significant difference in the M$_{\text{molgas}}$ (Fig. \ref{fig:mmol_hist}) and f$_{\text{mol}}$ distributions. Moreover, binning in M$_*$ and SFR, we found that the median f$_{\text{mol}}$ in each bin is practically the same for both samples (with the exception of the lowest M$_*$ and highest SFR bin, which have only one \nev{} source within). We also compared the mean host-galaxy IR SEDs and found that the two samples have similar SEDs at $\lambda_{\rm{rest}}>300\,\mu\rm{m}$ (thus explaining the similar amount of cold gas content), while the \nev\ sample has, on average, higher mid-IR emission (i.e., higher SFR). This is linked to the lower fraction of ``quiescent'' galaxies in the \nev\ sample, which have lower mid-IR emission with respect to main-sequence SF-galaxies. The different shapes of the SED in the mid-IR regime are linked to the fact that the \nev\ sample has, according to the SED-fitting results, a higher fraction of its dust luminosity coming from the hot ($130-250\,\rm{K}$) and warm ($30-60\,\rm{K}$) dust phases, namely, a higher emission in the $\lambda_{\rm{rest}}<100\mu\rm{m}$. \par
We conclude that the highest fraction of sources within the MS for the \nev{} sample and the lack of a population of ``quiescent'' galaxies is not related to a different amount of cold gas reservoir in the host galaxy, rather it is linked to a different efficiency in forming stars. We propose three explanations for this higher SF efficiency: the AGN may have the effect of enhancing the SF; the AGN is more likely to be triggered in galaxies with higher SFR; or the \nev\ selection efficiently allows us to pick up AGNs in an obscured growth phase, where we expect high SFR and obscured AGN activity. This last hypothesis is consistent with the \nev{} sample being slightly younger on average: the IR host galaxies are still growing and will reach higher masses than the control sample. While the physical explanation of the \nev-obscured AGN selection efficiency still needs to be fully understood, we speculate that the explanation of why the \nev\ selection may be well suited to select AGNs in the obscured accretion phase resides in the fact that its emission is effectively extinguished by even a modest quantity of dust in their host-galaxy. Thus, sources in the starburst phase (prior to the obscured accretion phase) are missed due to the (perceived) lack of \nev\ emission, while galaxies after the obscured accretion phase (thus transitioning to a type 1 AGN) are excluded due to our selection of objects with only narrow lines in their spectra.

\section{Interpretation within the in situ co-evolution model}\label{sec:insitu}
We place our sources in the context of the AGN-galaxy co-evolution paradigm by comparing them with the theoretical prediction of the in situ BH-galaxy evolution model \citep{mancuso16b, mancuso16a, mancuso17, lapi18}. This model allows us to easily follow the galaxy evolution, thus interpreting the properties of the \nev\ sources in the context of the AGN-galaxy co-evolution.\par
In the in situ co-evolution model, the SF is a local process regulated by energy feedback from supernova explosions (SNe) and from the central SMBH. In the first phase of the galaxy evolution, the dominant feedback is due to SNe, with SFR $\propto t^{1/2}$ and $M_* \propto t^{2/3}$. The MS emerges as the locus in the SFR$-M*$ plane where galaxies spend most of their time, thus where it is most probable to find them. In these early stages, the BH mass grows exponentially in gas-rich obscured environment, the AGN emits at mildly super Eddington ratio and its luminosity rises exponentially. After less than one Gyr (for massive galaxies), the AGN luminosity reaches values similar to those of the SF and becomes dominant. Via powerful winds and outflows, the AGN power heats up and removes the gas from the host galaxy, quenching the SF. The galaxy moves below the MS and the stellar populations evolve passively, while the residual gas in the central region accretes into the SMBH at lower sub-Eddington ratios and the AGN luminosity exponentially declines. At the end of this phase, what is left is a "red and dead" elliptical galaxy. \par
\begin{figure}
  \centering
  \resizebox{\hsize}{!}{\includegraphics{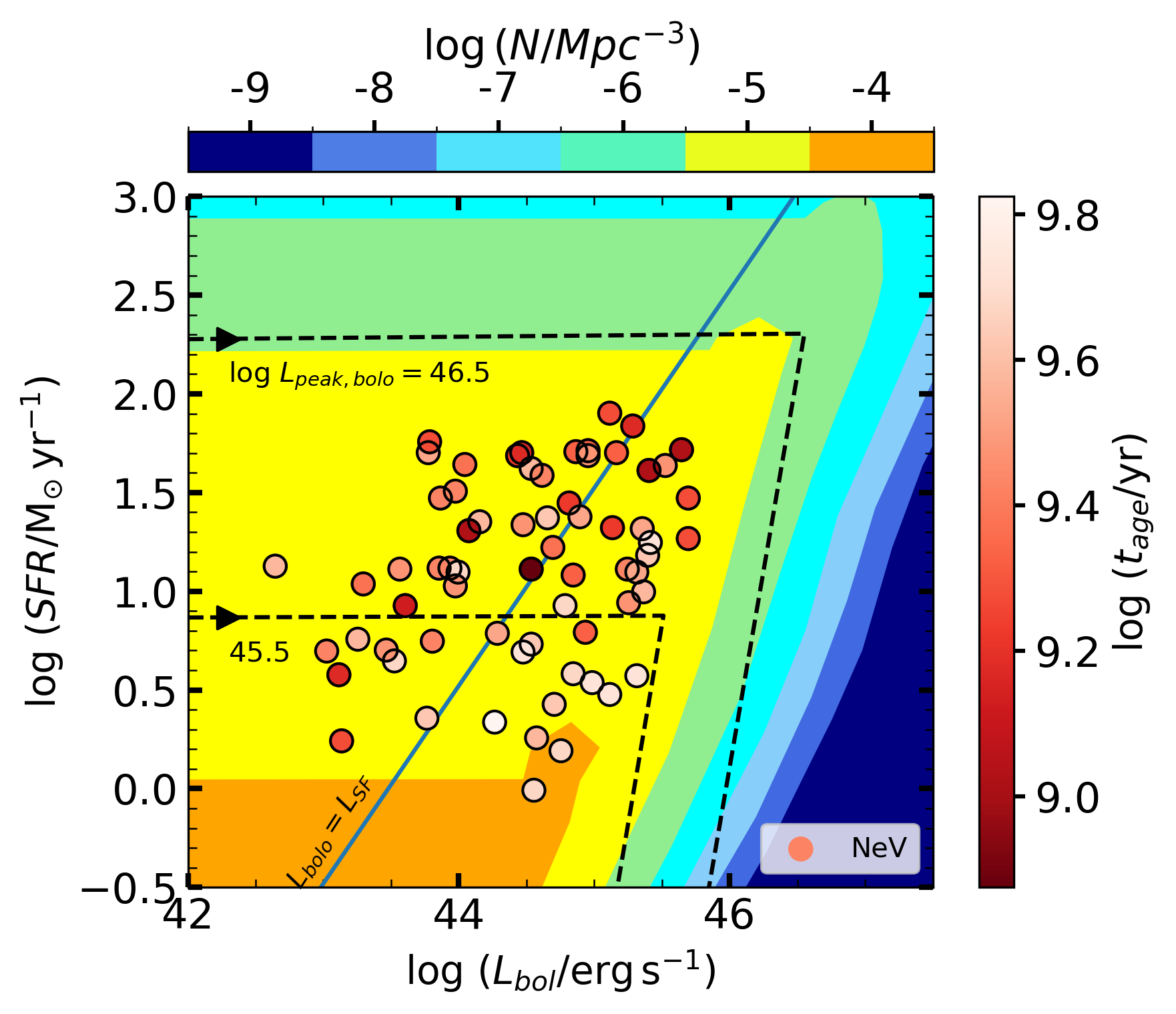}}
  \caption{Distribution of the \nev{} sources on the SFR$-$L$_{bol}$ plane. The color code indicates the age of the oldest stars in the host galaxy, as provided by \textsl{sed3fit}. Colored contours illustrate the number density of galaxies plus AGN at $z \sim 1$ as predicted by the in situ co-evolution model \citep{mancuso16b}; orange, yellow, green, cyan, light blue, blue, and dark blue contours refer to number density of $10^{-4}$, $10^{-5}$, $10^{-6}$, $10^{-7}$, $10^{-8}$, $10^{-9}$ Mpc$^{-1}$, respectively. The black dashed lines show two evolutionary tracks (forward time direction indicated by the arrows) for AGNs with peak bolometric luminosity of $10^{45.5}$ and $10^{46.5}$ erg$/$s. The blue continuous line indicates where the SF luminosity is equal to the AGN luminosity. The average uncertainty on the $t_{\rm{age}}$ is $\sim0.1\,$dex.}
  \label{fig:insitu}
\end{figure}

In Figure \ref{fig:insitu} we illustrate the position of the \nev{} sources in the SFR$-L_{bol}$ diagram, along with the prediction of the in situ model \citep{mancuso16b}. The \nev\ AGNs are color-coded on the basis of their t$_{\rm{age}}$ (which have a mean uncertainties of $\sim 0.1$ dex). The colored contours show the number density of sources predicted by the model. The dashed lines represent two evolutionary tracks from the in situ model for AGNs with different peak luminosity (the maximum bolometric luminosity an AGN can reach before the feedback kicks in and lowers the AGN output), with the forward time direction indicated by the arrows. The evolution of the sources begins from the left, with the SFR roughly constant while the AGN luminosity grows exponentially. When the source crosses the blue $L_{bol}=L_{SF}$ line, the AGN luminosity becomes dominant. In these two phases, the AGN accretes at Eddington and super Eddington ratios in obscured condition. Then, when the feedback starts to "kick in," the SF is abruptly quenched and the SFR drops quickly, while the AGN accretes the remaining material at lower sub-Eddington ratios and its emission fades with a slower trend.\par
According to the in situ prediction, our sources on the left of the blue line would be obscured and accreting AGNs, in which the SF is still dominated by stellar feedback. The sources on the right would be those for which the AGN is the main actor in the feedback processes. Finally, we note that there is a segregation of older sources in the lower right locus of the plot. These sources could be those for which the AGN has started to quench the SF and that are rapidly moving downward in the plot.\par
Therefore, according to the in situ model prediction, the \nev{} sources are located in the SFR$-L_{bol}$ plane where we expect AGNs in the obscured accretion phase, with the oldest sources being compatible to be at the beginning of the "quenching phase."

\section{Conclusion}\label{sec:conclusion}

In this work, we investigated the AGN and host galaxy properties of a sample of 94 \nev-selected type 2 AGNs. We performed an X-ray spectral analysis of the 36 X-ray-detected sources to characterize their AGN intrinsic luminosity and obscuration. For the X-ray-undetected sources, we used the X/\nev{} ratio to estimate their amount of obscuration and the fraction of CT-AGN. We performed optical-to-FIR SED-fitting, using the \textit{sed3fit} algorithm, to characterize both the AGN and the host galaxy properties. We used the stellar mass to obtain the BH mass and, subsequently, the Eddington ratio of the \nev{} sample. Finally, we compared the host galaxy properties (stellar mass, SFR, sSFR, cold gas content) of the \nev{} sample with those of a non-active control sample and interpreted our results in the light of the in situ BH-galaxy evolution scenario. Our main results are the following:
\begin{enumerate}
\item The \nev{} selection is an optimal tool to select very obscured ($N_H \geq 10^{23}$ cm$^{-2}$) and CT-AGNs at $z\sim 1$. More than two-thirds of our sample is composed of very obscured sources and $\sim 20\%$ of the sources are candidate CT-AGNs. 
\item Almost half of the sample is composed of AGNs in a strong episode of SMBH growth ($\lambda_{\text{Edd}}\ge 0.1$).
\item The \nev{} sample has a significantly higher fraction of sources within the MS than normal galaxies and seems to lack a population of "quiescent" (low sSFR) galaxies.
\item This difference is not due to the amount of cold gas reservoir, as both samples show similar M$_{\text{mol}}$, but to the higher efficiency in forming stars of the \nev{} sample. This higher efficiency could be related to the AGN enhancing the SF, to the high SF triggering the AGN activity, or it could be due to a selection effect: with the \nev{}-selection picking-up AGNs in their obscured growing phase.
\item We interpret our results in the context of the in situ co-evolution scenario and find that the \nev{} sample is compatible to be composed by sources in the "pre-quenching" phase, with only the few oldest sources showing the effect of the AGN quenching the SF. Therefore we favor the latter hypothesis: \nev\ sources are preferentially hosted in obscured AGNs that are efficiently accreting and forming stars. 
\end{enumerate} 
Multiwavelength analyses similar to this one, carried out using different high-ionization lines as selection method (i.e., \oiii\ for lower $z$, \civ\ for higher $z$), should allow us to study the redshift evolution of the AGN and host galaxy properties, as well as to unveil the mystery of the BH-galaxy co-evolution mechanism.
 
\begin{acknowledgements}
The authors would like to thank the anonymous referee for her/his comments and useful suggestions. R.G. acknowledges financial contribution from the agreement ASI-INAF n. 2017-14-H.O. A.L. is partly supported by the PRIN MIUR 2017 prot. 20173ML3WW 002 ``Opening the ALMA window on the cosmic evolution of gas, stars, and massive black holes''. C.V. and F.R. acknowledge support from PRIN MIUR 2017 PH3WAT (``Black hole winds and the baryon life cycle of galaxies'').
\end{acknowledgements}

\bibliographystyle{aa.bst} 
\bibliography{45288corr.bib} 

\begin{thebibliography}{122}
\expandafter\ifx\csname natexlab\endcsname\relax\def\natexlab#1{#1}\fi

\bibitem[{{Alexander} {et~al.}(2005){Alexander}, {Bauer}, {Chapman}, {Smail},
  {Blain}, {Brandt}, \& {Ivison}}]{alexander05}
{Alexander}, D.~M., {Bauer}, F.~E., {Chapman}, S.~C., {et~al.} 2005, \apj, 632,
  736

\bibitem[{{Alexander} {et~al.}(2010){Alexander}, {Swinbank}, {Smail},
  {McDermid}, \& {Nesvadba}}]{alexander10}
{Alexander}, D.~M., {Swinbank}, A.~M., {Smail}, I., {McDermid}, R., \&
  {Nesvadba}, N.~P.~H. 2010, \mnras, 402, 2211

\bibitem[{{Almaini}(2003)}]{almaini03}
{Almaini}, O. 2003, Astronomische Nachrichten, 324, 109

\bibitem[{{Ananna} {et~al.}(2019){Ananna}, {Treister}, {Urry}, {Ricci},
  {Kirkpatrick}, {LaMassa}, {Buchner}, {Civano}, {Tremmel}, \&
  {Marchesi}}]{ananna19}
{Ananna}, T.~T., {Treister}, E., {Urry}, C.~M., {et~al.} 2019, \apj, 871, 240

\bibitem[{{Archibald} {et~al.}(2002){Archibald}, {Dunlop}, {Jimenez},
  {Fria{\c{c}}a}, {McLure}, \& {Hughes}}]{archibald02}
{Archibald}, E.~N., {Dunlop}, J.~S., {Jimenez}, R., {et~al.} 2002, \mnras, 336,
  353

\bibitem[{{Arnaud}(1996)}]{xspec}
{Arnaud}, K.~A. 1996, in Astronomical Society of the Pacific Conference Series,
  Vol. 101, Astronomical Data Analysis Software and Systems V, ed. G.~H.
  {Jacoby} \& J.~{Barnes}, 17

\bibitem[{{Barchiesi} {et~al.}(2021){Barchiesi}, {Pozzi}, {Vignali}, {Carrera},
  {Vito}, {Calura}, {Bisigello}, {Lanzuisi}, {Gruppioni}, {Lusso},
  {Delvecchio}, {Negrello}, {Cooray}, {Feltre}, {Fern{\'a}ndez-Ontiveros},
  {Gallerani}, {Kaneda}, {Oyabu}, {Pereira-Santaella}, {Piconcelli}, {Ricci},
  {Rodighiero}, {Spinoglio}, \& {Tombesi}}]{barchiesi21_spica}
{Barchiesi}, L., {Pozzi}, F., {Vignali}, C., {et~al.} 2021, \pasa, 38, e033

\bibitem[{{Behar} {et~al.}(2015){Behar}, {Baldi}, {Laor}, {Horesh}, {Stevens},
  \& {Tzioumis}}]{behar15}
{Behar}, E., {Baldi}, R.~D., {Laor}, A., {et~al.} 2015, \mnras, 451, 517

\bibitem[{{Bernhard} {et~al.}(2018){Bernhard}, {Mullaney}, {Aird}, {Hickox},
  {Jones}, {Stanley}, {Grimmett}, \& {Daddi}}]{bernhard18}
{Bernhard}, E., {Mullaney}, J.~R., {Aird}, J., {et~al.} 2018, \mnras, 476, 436

\bibitem[{{Bernhard} {et~al.}(2016){Bernhard}, {Mullaney}, {Daddi}, {Ciesla},
  \& {Schreiber}}]{bernhard16}
{Bernhard}, E., {Mullaney}, J.~R., {Daddi}, E., {Ciesla}, L., \& {Schreiber},
  C. 2016, \mnras, 460, 902

\bibitem[{{Berta} {et~al.}(2013){Berta}, {Lutz}, {Santini}, {Wuyts}, {Rosario},
  {Brisbin}, {Cooray}, {Franceschini}, {Gruppioni}, {Hatziminaoglou}, {Hwang},
  {Le Floc'h}, {Magnelli}, {Nordon}, {Oliver}, {Page}, {Popesso}, {Pozzetti},
  {Pozzi}, {Riguccini}, {Rodighiero}, {Roseboom}, {Scott}, {Symeonidis},
  {Valtchanov}, {Viero}, \& {Wang}}]{berta13}
{Berta}, S., {Lutz}, D., {Santini}, P., {et~al.} 2013, \aap, 551, A100

\bibitem[{{Bevington} \& {Robinson}(2003)}]{bevington03}
{Bevington}, P.~R. \& {Robinson}, D.~K. 2003, {Data reduction and error
  analysis for the physical sciences}

\bibitem[{{Bisigello} {et~al.}(2021){Bisigello}, {Gruppioni}, {Feltre},
  {Calura}, {Pozzi}, {Vignali}, {Barchiesi}, {Rodighiero}, \&
  {Negrello}}]{bisigello21a}
{Bisigello}, L., {Gruppioni}, C., {Feltre}, A., {et~al.} 2021, \aap, 651, A52

\bibitem[{{Boquien} {et~al.}(2019){Boquien}, {Burgarella}, {Roehlly}, {Buat},
  {Ciesla}, {Corre}, {Inoue}, \& {Salas}}]{boquien19}
{Boquien}, M., {Burgarella}, D., {Roehlly}, Y., {et~al.} 2019, \aap, 622, A103

\bibitem[{Bradford {et~al.}(2022)Bradford, Glenn, Armus, Battersby, Bolatto,
  Hensley, Meixner, Mills, Pontoppidan, Pope, Smith, Staguhn, \&
  Groups}]{bradford22}
Bradford, C., Glenn, J., Armus, L., {et~al.} 2022, Bulletin of the AAS, 54,
  https://baas.aas.org/pub/2022n6i304p07

\bibitem[{{Brusa} {et~al.}(2015){Brusa}, {Bongiorno}, {Cresci}, {Perna},
  {Marconi}, {Mainieri}, {Maiolino}, {Salvato}, {Lusso}, {Santini}, {Comastri},
  {Fiore}, {Gilli}, {La Franca}, {Lanzuisi}, {Lutz}, {Merloni}, {Mignoli},
  {Onori}, {Piconcelli}, {Rosario}, {Vignali}, \& {Zamorani}}]{brusa15}
{Brusa}, M., {Bongiorno}, A., {Cresci}, G., {et~al.} 2015, \mnras, 446, 2394

\bibitem[{{Brusa} {et~al.}(2016){Brusa}, {Perna}, {Cresci}, {Schramm},
  {Delvecchio}, {Lanzuisi}, {Mainieri}, {Mignoli}, {Zamorani}, {Berta},
  {Bongiorno}, {Comastri}, {Fiore}, {Kakkad}, {Marconi}, {Rosario}, {Contini},
  \& {Lamareille}}]{brusa16}
{Brusa}, M., {Perna}, M., {Cresci}, G., {et~al.} 2016, \aap, 588, A58

\bibitem[{{Bruzual}(2007)}]{bruzual07}
{Bruzual}, G. 2007, in Astronomical Society of the Pacific Conference Series,
  Vol. 374, From Stars to Galaxies: Building the Pieces to Build Up the
  Universe, ed. A.~{Vallenari}, R.~{Tantalo}, L.~{Portinari}, \& A.~{Moretti},
  303

\bibitem[{{Buchner} {et~al.}(2015){Buchner}, {Georgakakis}, {Nandra},
  {Brightman}, {Menzel}, {Liu}, {Hsu}, {Salvato}, {Rangel}, {Aird}, {Merloni},
  \& {Ross}}]{buchenr15}
{Buchner}, J., {Georgakakis}, A., {Nandra}, K., {et~al.} 2015, \apj, 802, 89

\bibitem[{{Burgarella} {et~al.}(2005){Burgarella}, {Buat}, \&
  {Iglesias-P{\'a}ramo}}]{burgarella05}
{Burgarella}, D., {Buat}, V., \& {Iglesias-P{\'a}ramo}, J. 2005, \mnras, 360,
  1413

\bibitem[{{Cappi} {et~al.}(2006){Cappi}, {Panessa}, {Bassani}, {Dadina}, {Di
  Cocco}, {Comastri}, {della Ceca}, {Filippenko}, {Gianotti}, {Ho}, {Malaguti},
  {Mulchaey}, {Palumbo}, {Piconcelli}, {Sargent}, {Stephen}, {Trifoglio}, \&
  {Weaver}}]{cappi06}
{Cappi}, M., {Panessa}, F., {Bassani}, L., {et~al.} 2006, \aap, 446, 459

\bibitem[{{Cash}(1979)}]{cash79}
{Cash}, W. 1979, \apj, 228, 939

\bibitem[{{Cattaneo} {et~al.}(2009){Cattaneo}, {Faber}, {Binney}, {Dekel},
  {Kormendy}, {Mushotzky}, {Babul}, {Best}, {Br{\"u}ggen}, {Fabian}, {Frenk},
  {Khalatyan}, {Netzer}, {Mahdavi}, {Silk}, {Steinmetz}, \&
  {Wisotzki}}]{cattaneo09}
{Cattaneo}, A., {Faber}, S.~M., {Binney}, J., {et~al.} 2009, \nat, 460, 213

\bibitem[{{Chabrier}(2003)}]{chabrier03}
{Chabrier}, G. 2003, Publications of the Astronomical Society of the Pacific,
  115, 763

\bibitem[{{Charlot} \& {Fall}(2000)}]{charlot00}
{Charlot}, S. \& {Fall}, S.~M. 2000, \apj, 539, 718

\bibitem[{{Cicone} {et~al.}(2014){Cicone}, {Maiolino}, {Sturm},
  {Graci{\'a}-Carpio}, {Feruglio}, {Neri}, {Aalto}, {Davies}, {Fiore},
  {Fischer}, {Garc{\'\i}a-Burillo}, {Gonz{\'a}lez-Alfonso}, {Hailey-Dunsheath},
  {Piconcelli}, \& {Veilleux}}]{cicone14}
{Cicone}, C., {Maiolino}, R., {Sturm}, E., {et~al.} 2014, \aap, 562, A21

\bibitem[{{Civano} {et~al.}(2016){Civano}, {Marchesi}, {Comastri}, {Urry},
  {Elvis}, {Cappelluti}, {Puccetti}, {Brusa}, {Zamorani}, {Hasinger},
  {Aldcroft}, {Alexand er}, {Allevato}, {Brunner}, {Capak}, {Finoguenov},
  {Fiore}, {Fruscione}, {Gilli}, {Glotfelty}, {Griffiths}, {Hao}, {Harrison},
  {Jahnke}, {Kartaltepe}, {Karim}, {LaMassa}, {Lanzuisi}, {Miyaji}, {Ranalli},
  {Salvato}, {Sargent}, {Scoville}, {Schawinski}, {Schinnerer}, {Silverman},
  {Smolcic}, {Stern}, {Toft}, {Trakhtenbrot}, {Treister}, \&
  {Vignali}}]{civano16}
{Civano}, F., {Marchesi}, S., {Comastri}, A., {et~al.} 2016, \apj, 819, 62

\bibitem[{{Cleri} {et~al.}(2022){Cleri}, {Yang}, {Papovich}, {Trump},
  {Backhaus}, {Estrada-Carpenter}, {Finkelstein}, {Giavalisco}, {Hutchison},
  {Ji}, {Jung}, {Matharu}, {Momcheva}, {Olivier}, {Simons}, \&
  {Weiner}}]{cleri22}
{Cleri}, N.~J., {Yang}, G., {Papovich}, C., {et~al.} 2022, arXiv e-prints,
  arXiv:2209.06247

\bibitem[{{Combes} {et~al.}(2019){Combes}, {Garc{\'\i}a-Burillo}, {Audibert},
  {Hunt}, {Eckart}, {Aalto}, {Casasola}, {Boone}, {Krips}, {Viti}, {Sakamoto},
  {Muller}, {Dasyra}, {van der Werf}, \& {Martin}}]{combes19}
{Combes}, F., {Garc{\'\i}a-Burillo}, S., {Audibert}, A., {et~al.} 2019,
  Astronomy and Astrophysics, 623, A79

\bibitem[{{da Cunha} {et~al.}(2008){da Cunha}, {Charlot}, \& {Elbaz}}]{cuna08}
{da Cunha}, E., {Charlot}, S., \& {Elbaz}, D. 2008, \mnras, 388, 1595

\bibitem[{{Delvecchio} {et~al.}(2014){Delvecchio}, {Gruppioni}, {Pozzi},
  {Berta}, {Zamorani}, {Cimatti}, {Lutz}, {Scott}, {Vignali}, {Cresci},
  {Feltre}, {Cooray}, {Vaccari}, {Fritz}, {Le Floc'h}, {Magnelli}, {Popesso},
  {Oliver}, {Bock}, {Carollo}, {Contini}, {Le F{\'e}vre}, {Lilly}, {Mainieri},
  {Renzini}, \& {Scodeggio}}]{delvecchio14}
{Delvecchio}, I., {Gruppioni}, C., {Pozzi}, F., {et~al.} 2014, \mnras, 439,
  2736

\bibitem[{{Dessauges-Zavadsky} {et~al.}(2020){Dessauges-Zavadsky}, {Ginolfi},
  {Pozzi}, {B{\'e}thermin}, {Le F{\`e}vre}, {Fujimoto}, {Silverman}, {Jones},
  {Vallini}, {Schaerer}, {Faisst}, {Khusanova}, {Fudamoto}, {Cassata},
  {Loiacono}, {Capak}, {Yan}, {Amorin}, {Bardelli}, {Boquien}, {Cimatti},
  {Gruppioni}, {Hathi}, {Ibar}, {Koekemoer}, {Lemaux}, {Narayanan}, {Oesch},
  {Rodighiero}, {Romano}, {Talia}, {Toft}, {Vergani}, {Zamorani}, \&
  {Zucca}}]{dessauges20}
{Dessauges-Zavadsky}, M., {Ginolfi}, M., {Pozzi}, F., {et~al.} 2020, \aap, 643,
  A5

\bibitem[{{Di Matteo} {et~al.}(2005){Di Matteo}, {Springel}, \&
  {Hernquist}}]{dimatteo05}
{Di Matteo}, T., {Springel}, V., \& {Hernquist}, L. 2005, \nat, 433, 604

\bibitem[{{Duras} {et~al.}(2020){Duras}, {Bongiorno}, {Ricci}, {Piconcelli},
  {Shankar}, {Lusso}, {Bianchi}, {Fiore}, {Maiolino}, {Marconi}, {Onori},
  {Sani}, {Schneider}, {Vignali}, \& {La Franca}}]{duras20}
{Duras}, F., {Bongiorno}, A., {Ricci}, F., {et~al.} 2020, \aap, 636, A73

\bibitem[{{Elitzur}(2008)}]{elitzur08}
{Elitzur}, M. 2008, New Astronomy Reviews, 52, 274

\bibitem[{{Ellison} {et~al.}(2019){Ellison}, {Viswanathan}, {Patton},
  {Bottrell}, {McConnachie}, {Gwyn}, \& {Cuillandre}}]{ellison19}
{Ellison}, S.~L., {Viswanathan}, A., {Patton}, D.~R., {et~al.} 2019, \mnras,
  487, 2491

\bibitem[{{Elvis} {et~al.}(2009){Elvis}, {Civano}, {Vignali}, {Puccetti},
  {Fiore}, {Cappelluti}, {Aldcroft}, {Fruscione}, {Zamorani}, {Comastri},
  {Brusa}, {Gilli}, {Miyaji}, {Damiani}, {Koekemoer}, {Finoguenov}, {Brunner},
  {Urry}, {Silverman}, {Mainieri}, {Hasinger}, {Griffiths}, {Carollo}, {Hao},
  {Guzzo}, {Blain}, {Calzetti}, {Carilli}, {Capak}, {Ettori}, {Fabbiano},
  {Impey}, {Lilly}, {Mobasher}, {Rich}, {Salvato}, {Sand ers}, {Schinnerer},
  {Scoville}, {Shopbell}, {Taylor}, {Taniguchi}, \& {Volonteri}}]{elvis09}
{Elvis}, M., {Civano}, F., {Vignali}, C., {et~al.} 2009, \apjs, 184, 158

\bibitem[{{Feltre} {et~al.}(2012){Feltre}, {Hatziminaoglou}, {Fritz}, \&
  {Franceschini}}]{feltre12}
{Feltre}, A., {Hatziminaoglou}, E., {Fritz}, J., \& {Franceschini}, A. 2012,
  \mnras, 426, 120

\bibitem[{{Ferrarese} \& {Merritt}(2000)}]{ferrarese00}
{Ferrarese}, L. \& {Merritt}, D. 2000, \apjl, 539, L9

\bibitem[{{Feruglio} {et~al.}(2010){Feruglio}, {Maiolino}, {Piconcelli},
  {Menci}, {Aussel}, {Lamastra}, \& {Fiore}}]{feruglio10}
{Feruglio}, C., {Maiolino}, R., {Piconcelli}, E., {et~al.} 2010, \aap, 518,
  L155

\bibitem[{{Fritz} {et~al.}(2006){Fritz}, {Franceschini}, \&
  {Hatziminaoglou}}]{fritz06}
{Fritz}, J., {Franceschini}, A., \& {Hatziminaoglou}, E. 2006, \mnras, 366, 767

\bibitem[{{Fruscione} {et~al.}(2006){Fruscione}, {McDowell}, {Allen},
  {Brickhouse}, {Burke}, {Davis}, {Durham}, {Elvis}, {Galle}, {Harris},
  {Huenemoerder}, {Houck}, {Ishibashi}, {Karovska}, {Nicastro}, {Noble},
  {Nowak}, {Primini}, {Siemiginowska}, {Smith}, \& {Wise}}]{ciao}
{Fruscione}, A., {McDowell}, J.~C., {Allen}, G.~E., {et~al.} 2006, in Society
  of Photo-Optical Instrumentation Engineers (SPIE) Conference Series, Vol.
  6270, \procspie, 62701V

\bibitem[{{Gebhardt} {et~al.}(2000){Gebhardt}, {Bender}, {Bower}, {Dressler},
  {Faber}, {Filippenko}, {Green}, {Grillmair}, {Ho}, {Kormendy}, {Lauer},
  {Magorrian}, {Pinkney}, {Richstone}, \& {Tremaine}}]{gebhardt00}
{Gebhardt}, K., {Bender}, R., {Bower}, G., {et~al.} 2000, \apjl, 539, L13

\bibitem[{{Gehrels}(1986)}]{gehrels86}
{Gehrels}, N. 1986, \apj, 303, 336

\bibitem[{{Gilli}(2013)}]{gilli13}
{Gilli}, R. 2013, \memsai, 84, 647

\bibitem[{{Gilli} {et~al.}(2007){Gilli}, {Comastri}, \& {Hasinger}}]{gilli07}
{Gilli}, R., {Comastri}, A., \& {Hasinger}, G. 2007, \aap, 463, 79

\bibitem[{{Gilli} {et~al.}(2010){Gilli}, {Vignali}, {Mignoli}, {Iwasawa},
  {Comastri}, \& {Zamorani}}]{gilli10}
{Gilli}, R., {Vignali}, C., {Mignoli}, M., {et~al.} 2010, \aap, 519, A92

\bibitem[{{Glikman} {et~al.}(2012){Glikman}, {Urrutia}, {Lacy}, {Djorgovski},
  {Mahabal}, {Myers}, {Ross}, {Petitjean}, {Ge}, {Schneider}, \&
  {York}}]{glikman12}
{Glikman}, E., {Urrutia}, T., {Lacy}, M., {et~al.} 2012, \apj, 757, 51

\bibitem[{{Gruppioni} {et~al.}(2016){Gruppioni}, {Berta}, {Spinoglio},
  {Pereira-Santaella}, {Pozzi}, {Andreani}, {Bonato}, {De Zotti}, {Malkan},
  {Negrello}, {Vallini}, \& {Vignali}}]{gruppioni16}
{Gruppioni}, C., {Berta}, S., {Spinoglio}, L., {et~al.} 2016, \mnras, 458, 4297

\bibitem[{{Harikane} {et~al.}(2023){Harikane}, {Zhang}, {Nakajima}, {Ouchi},
  {Isobe}, {Ono}, {Hatano}, {Xu}, \& {Umeda}}]{harikane23}
{Harikane}, Y., {Zhang}, Y., {Nakajima}, K., {et~al.} 2023, \apj, 959, 39

\bibitem[{{Harrison} {et~al.}(2012){Harrison}, {Alexander}, {Swinbank},
  {Smail}, {Alaghband-Zadeh}, {Bauer}, {Chapman}, {Del Moro}, {Hickox},
  {Ivison}, {Men{\'e}ndez-Delmestre}, {Mullaney}, \& {Nesvadba}}]{harrison12}
{Harrison}, C.~M., {Alexander}, D.~M., {Swinbank}, A.~M., {et~al.} 2012,
  \mnras, 426, 1073

\bibitem[{{Harrison} {et~al.}(2018){Harrison}, {Costa}, {Tadhunter},
  {Fl{\"u}tsch}, {Kakkad}, {Perna}, \& {Vietri}}]{harrison18}
{Harrison}, C.~M., {Costa}, T., {Tadhunter}, C.~N., {et~al.} 2018, Nature
  Astronomy, 2, 198

\bibitem[{{Hopkins} {et~al.}(2007{\natexlab{a}}){Hopkins}, {Lidz}, {Hernquist},
  {Coil}, {Myers}, {Cox}, \& {Spergel}}]{hopkins08}
{Hopkins}, P.~F., {Lidz}, A., {Hernquist}, L., {et~al.} 2007{\natexlab{a}},
  \apj, 662, 110

\bibitem[{{Hopkins} {et~al.}(2007{\natexlab{b}}){Hopkins}, {Richards}, \&
  {Hernquist}}]{hopkins07}
{Hopkins}, P.~F., {Richards}, G.~T., \& {Hernquist}, L. 2007{\natexlab{b}},
  \apj, 654, 731

\bibitem[{{Iwasawa} {et~al.}(2020){Iwasawa}, {Comastri}, {Vignali}, {Gilli},
  {Lanzuisi}, {Brandt}, {Tozzi}, {Brusa}, {Carrera}, {Ranalli}, {Mainieri},
  {Georgantopoulos}, {Puccetti}, \& {Paolillo}}]{iwasawa20}
{Iwasawa}, K., {Comastri}, A., {Vignali}, C., {et~al.} 2020, \aap, 639, A51

\bibitem[{{Jaffe} {et~al.}(2004){Jaffe}, {Meisenheimer}, {R{\"o}ttgering},
  {Leinert}, {Richichi}, {Chesneau}, {Fraix-Burnet}, {Glazenborg-Kluttig},
  {Granato}, {Graser}, {Heijligers}, {K{\"o}hler}, {Malbet}, {Miley},
  {Paresce}, {Pel}, {Perrin}, {Przygodda}, {Schoeller}, {Sol}, {Waters},
  {Weigelt}, {Woillez}, \& {de Zeeuw}}]{jaffe04}
{Jaffe}, W., {Meisenheimer}, K., {R{\"o}ttgering}, H.~J.~A., {et~al.} 2004,
  \nat, 429, 47

\bibitem[{{Juneau} {et~al.}(2011){Juneau}, {Dickinson}, {Alexander}, \&
  {Salim}}]{juneau11}
{Juneau}, S., {Dickinson}, M., {Alexander}, D.~M., \& {Salim}, S. 2011, \apj,
  736, 104

\bibitem[{{Kaasinen} {et~al.}(2019){Kaasinen}, {Scoville}, {Walter}, {Da
  Cunha}, {Popping}, {Pavesi}, {Darvish}, {Casey}, {Riechers}, \&
  {Glover}}]{kaasinen19}
{Kaasinen}, M., {Scoville}, N., {Walter}, F., {et~al.} 2019, \apj, 880, 15

\bibitem[{{Kakkad} {et~al.}(2016){Kakkad}, {Mainieri}, {Padovani}, {Cresci},
  {Husemann}, {Carniani}, {Brusa}, {Lamastra}, {Lanzuisi}, {Piconcelli}, \&
  {Schramm}}]{kakkad16}
{Kakkad}, D., {Mainieri}, V., {Padovani}, P., {et~al.} 2016, \aap, 592, A148

\bibitem[{{Kalberla} {et~al.}(2005){Kalberla}, {Burton}, {Hartmann}, {Arnal},
  {Bajaja}, {Morras}, \& {P{\"o}ppel}}]{kalberla05}
{Kalberla}, P.~M.~W., {Burton}, W.~B., {Hartmann}, D., {et~al.} 2005, \aap,
  440, 775

\bibitem[{{Kawamuro} {et~al.}(2022){Kawamuro}, {Ricci}, {Imanishi},
  {Mushotzky}, {Izumi}, {Ricci}, {Bauer}, {Koss}, {Trakhtenbrot}, {Ichikawa},
  {Rojas}, {Smith}, {Shimizu}, {Oh}, {den Brok}, {Baba}, {Balokovi{\'c}},
  {Chang}, {Kakkad}, {Pfeifle}, {Privon}, {Temple}, {Ueda}, {Harrison},
  {Powell}, {Stern}, {Urry}, \& {Sanders}}]{kawamuro22}
{Kawamuro}, T., {Ricci}, C., {Imanishi}, M., {et~al.} 2022, arXiv e-prints,
  arXiv:2208.03880

\bibitem[{{Kennicutt}(1998)}]{kennicutt98}
{Kennicutt}, Robert~C., J. 1998, \araa, 36, 189

\bibitem[{{Kormendy} \& {Ho}(2013)}]{kormendy13}
{Kormendy}, J. \& {Ho}, L.~C. 2013, \araa, 51, 511

\bibitem[{{Kormendy} \& {Richstone}(1995)}]{kormendy95}
{Kormendy}, J. \& {Richstone}, D. 1995, \araa, 33, 581

\bibitem[{{Laigle} {et~al.}(2016){Laigle}, {McCracken}, {Ilbert}, {Hsieh},
  {Davidzon}, {Capak}, {Hasinger}, {Silverman}, {Pichon}, {Coupon}, {Aussel},
  {Le Borgne}, {Caputi}, {Cassata}, {Chang}, {Civano}, {Dunlop}, {Fynbo},
  {Kartaltepe}, {Koekemoer}, {Le F{\`e}vre}, {Le Floc'h}, {Leauthaud}, {Lilly},
  {Lin}, {Marchesi}, {Milvang-Jensen}, {Salvato}, {Sanders}, {Scoville},
  {Smolcic}, {Stockmann}, {Taniguchi}, {Tasca}, {Toft}, {Vaccari}, \&
  {Zabl}}]{laigle16}
{Laigle}, C., {McCracken}, H.~J., {Ilbert}, O., {et~al.} 2016, \apjs, 224, 24

\bibitem[{{LaMassa} {et~al.}(2017){LaMassa}, {Yaqoob}, \&
  {Kilgard}}]{lamassa17}
{LaMassa}, S.~M., {Yaqoob}, T., \& {Kilgard}, R. 2017, \apj, 840, 11

\bibitem[{{Lamastra} {et~al.}(2013){Lamastra}, {Menci}, {Fiore}, {Santini},
  {Bongiorno}, \& {Piconcelli}}]{lamastra13}
{Lamastra}, A., {Menci}, N., {Fiore}, F., {et~al.} 2013, \aap, 559, A56

\bibitem[{{Lanzuisi} {et~al.}(2018){Lanzuisi}, {Civano}, {Marchesi},
  {Comastri}, {Brusa}, {Gilli}, {Vignali}, {Zamorani}, {Brightman},
  {Griffiths}, \& {Koekemoer}}]{lanzuisi18}
{Lanzuisi}, G., {Civano}, F., {Marchesi}, S., {et~al.} 2018, \mnras, 480, 2578

\bibitem[{{Lapi} {et~al.}(2018){Lapi}, {Pantoni}, {Zanisi}, {Shi}, {Mancuso},
  {Massardi}, {Shankar}, {Bressan}, \& {Danese}}]{lapi18}
{Lapi}, A., {Pantoni}, L., {Zanisi}, L., {et~al.} 2018, \apj, 857, 22

\bibitem[{{Lapi} {et~al.}(2014){Lapi}, {Raimundo}, {Aversa}, {Cai}, {Negrello},
  {Celotti}, {De Zotti}, \& {Danese}}]{lapi14}
{Lapi}, A., {Raimundo}, S., {Aversa}, R., {et~al.} 2014, \apj, 782, 69

\bibitem[{{Lilly} {et~al.}(2009){Lilly}, {Le Brun}, {Maier}, {Mainieri},
  {Mignoli}, {Scodeggio}, {Zamorani}, {Carollo}, {Contini}, {Kneib}, {Le
  F{\`e}vre}, {Renzini}, {Bardelli}, {Bolzonella}, {Bongiorno}, {Caputi},
  {Coppa}, {Cucciati}, {de la Torre}, {de Ravel}, {Franzetti}, {Garilli},
  {Iovino}, {Kampczyk}, {Kovac}, {Knobel}, {Lamareille}, {Le Borgne}, {Pello},
  {Peng}, {P{\'e}rez-Montero}, {Ricciardelli}, {Silverman}, {Tanaka}, {Tasca},
  {Tresse}, {Vergani}, {Zucca}, {Ilbert}, {Salvato}, {Oesch}, {Abbas},
  {Bottini}, {Capak}, {Cappi}, {Cassata}, {Cimatti}, {Elvis}, {Fumana},
  {Guzzo}, {Hasinger}, {Koekemoer}, {Leauthaud}, {Maccagni}, {Marinoni},
  {McCracken}, {Memeo}, {Meneux}, {Porciani}, {Pozzetti}, {Sanders},
  {Scaramella}, {Scarlata}, {Scoville}, {Shopbell}, \& {Taniguchi}}]{lilly09}
{Lilly}, S.~J., {Le Brun}, V., {Maier}, C., {et~al.} 2009, \apjs, 184, 218

\bibitem[{{Lilly} {et~al.}(2007){Lilly}, {Le F{\`e}vre}, {Renzini}, {Zamorani},
  {Scodeggio}, {Contini}, {Carollo}, {Hasinger}, {Kneib}, {Iovino}, {Le Brun},
  {Maier}, {Mainieri}, {Mignoli}, {Silverman}, {Tasca}, {Bolzonella},
  {Bongiorno}, {Bottini}, {Capak}, {Caputi}, {Cimatti}, {Cucciati}, {Daddi},
  {Feldmann}, {Franzetti}, {Garilli}, {Guzzo}, {Ilbert}, {Kampczyk}, {Kovac},
  {Lamareille}, {Leauthaud}, {Le Borgne}, {McCracken}, {Marinoni}, {Pello},
  {Ricciardelli}, {Scarlata}, {Vergani}, {Sanders}, {Schinnerer}, {Scoville},
  {Taniguchi}, {Arnouts}, {Aussel}, {Bardelli}, {Brusa}, {Cappi}, {Ciliegi},
  {Finoguenov}, {Foucaud}, {Franceschini}, {Halliday}, {Impey}, {Knobel},
  {Koekemoer}, {Kurk}, {Maccagni}, {Maddox}, {Marano}, {Marconi}, {Meneux},
  {Mobasher}, {Moreau}, {Peacock}, {Porciani}, {Pozzetti}, {Scaramella},
  {Schiminovich}, {Shopbell}, {Smail}, {Thompson}, {Tresse}, {Vettolani},
  {Zanichelli}, \& {Zucca}}]{lilly07}
{Lilly}, S.~J., {Le F{\`e}vre}, O., {Renzini}, A., {et~al.} 2007, \apjs, 172,
  70

\bibitem[{{Lusso} {et~al.}(2012){Lusso}, {Comastri}, {Simmons}, {Mignoli},
  {Zamorani}, {Vignali}, {Brusa}, {Shankar}, {Lutz}, {Trump}, {Maiolino},
  {Gilli}, {Bolzonella}, {Puccetti}, {Salvato}, {Impey}, {Civano}, {Elvis},
  {Mainieri}, {Silverman}, {Koekemoer}, {Bongiorno}, {Merloni}, {Berta}, {Le
  Floc'h}, {Magnelli}, {Pozzi}, \& {Riguccini}}]{lusso12}
{Lusso}, E., {Comastri}, A., {Simmons}, B.~D., {et~al.} 2012, Monthly Notices
  of the Royal Astronomical Society, 425, 623

\bibitem[{{Magorrian} {et~al.}(1998){Magorrian}, {Tremaine}, {Richstone},
  {Bender}, {Bower}, {Dressler}, {Faber}, {Gebhardt}, {Green}, {Grillmair},
  {Kormendy}, \& {Lauer}}]{magorrian98}
{Magorrian}, J., {Tremaine}, S., {Richstone}, D., {et~al.} 1998, \aj, 115, 2285

\bibitem[{{Maiolino} {et~al.}(1998){Maiolino}, {Salvati}, {Bassani}, {Dadina},
  {della Ceca}, {Matt}, {Risaliti}, \& {Zamorani}}]{maiolino98}
{Maiolino}, R., {Salvati}, M., {Bassani}, L., {et~al.} 1998, \aap, 338, 781

\bibitem[{{Mancuso} {et~al.}(2017){Mancuso}, {Lapi}, {Prandoni}, {Obi},
  {Gonzalez-Nuevo}, {Perrotta}, {Bressan}, {Celotti}, \& {Danese}}]{mancuso17}
{Mancuso}, C., {Lapi}, A., {Prandoni}, I., {et~al.} 2017, \apj, 842, 95

\bibitem[{{Mancuso} {et~al.}(2016{\natexlab{a}}){Mancuso}, {Lapi}, {Shi},
  {Cai}, {Gonzalez-Nuevo}, {B{\'e}thermin}, \& {Danese}}]{mancuso16b}
{Mancuso}, C., {Lapi}, A., {Shi}, J., {et~al.} 2016{\natexlab{a}}, \apj, 833,
  152

\bibitem[{{Mancuso} {et~al.}(2016{\natexlab{b}}){Mancuso}, {Lapi}, {Shi},
  {Gonzalez-Nuevo}, {Aversa}, \& {Danese}}]{mancuso16a}
{Mancuso}, C., {Lapi}, A., {Shi}, J., {et~al.} 2016{\natexlab{b}}, \apj, 823,
  128

\bibitem[{{Marchesi} {et~al.}(2016){Marchesi}, {Lanzuisi}, {Civano}, {Iwasawa},
  {Suh}, {Comastri}, {Zamorani}, {Allevato}, {Griffiths}, {Miyaji}, {Ranalli},
  {Salvato}, {Schawinski}, {Silverman}, {Treister}, {Urry}, \&
  {Vignali}}]{marchesi16b}
{Marchesi}, S., {Lanzuisi}, G., {Civano}, F., {et~al.} 2016, \apj, 830, 100

\bibitem[{{Mignoli} {et~al.}(2019){Mignoli}, {Feltre}, {Bongiorno}, {Calura},
  {Gilli}, {Vignali}, {Zamorani}, {Lilly}, {Le F{\`e}vre}, {Bardelli},
  {Bolzonella}, {Bordoloi}, {Le Brun}, {Caputi}, {Cimatti}, {Diener},
  {Garilli}, {Koekemoer}, {Maier}, {Mainieri}, {Peng}, {P{\'e}rez Montero},
  {Silverman}, \& {Zucca}}]{mignoli19}
{Mignoli}, M., {Feltre}, A., {Bongiorno}, A., {et~al.} 2019, Astronomy and
  Astrophysics, 626, A9

\bibitem[{{Mignoli} {et~al.}(2013){Mignoli}, {Vignali}, {Gilli}, {Comastri},
  {Zamorani}, {Bolzonella}, {Bongiorno}, {Lamareille}, {Nair}, {Pozzetti},
  {Lilly}, {Carollo}, {Contini}, {Kneib}, {Le F{\`e}vre}, {Mainieri},
  {Renzini}, {Scodeggio}, {Bardelli}, {Caputi}, {Cucciati}, {de la Torre}, {de
  Ravel}, {Franzetti}, {Garilli}, {Iovino}, {Kampczyk}, {Knobel},
  {Kova{\v{c}}}, {Le Borgne}, {Le Brun}, {Maier}, {Pell{\`o}}, {Peng}, {Perez
  Montero}, {Presotto}, {Silverman}, {Tanaka}, {Tasca}, {Tresse}, {Vergani},
  {Zucca}, {Bordoloi}, {Cappi}, {Cimatti}, {Koekemoer}, {McCracken}, {Moresco},
  \& {Welikala}}]{mignoli13}
{Mignoli}, M., {Vignali}, C., {Gilli}, R., {et~al.} 2013, \aap, 556, A29

\bibitem[{{Moretti} {et~al.}(2012){Moretti}, {Vattakunnel}, {Tozzi},
  {Salvaterra}, {Severgnini}, {Fugazza}, {Haardt}, \& {Gilli}}]{moretti12}
{Moretti}, A., {Vattakunnel}, S., {Tozzi}, P., {et~al.} 2012, \aap, 548, A87

\bibitem[{{Nair} \& {Abraham}(2010)}]{nair2010}
{Nair}, P.~B. \& {Abraham}, R.~G. 2010, \apjs, 186, 427

\bibitem[{{Nesvadba} {et~al.}(2008){Nesvadba}, {Lehnert}, {De Breuck},
  {Downes}, {Neri}, {van Breugel}, {Walter}, {Kaiser}, {Binette}, \&
  {Kauffmann}}]{nesvadba08}
{Nesvadba}, N.~P.~H., {Lehnert}, M.~D., {De Breuck}, C., {et~al.} 2008, in Gas
  and Stars in Galaxies - A Multi-Wavelength 3D Perspective, 37

\bibitem[{{Noeske} {et~al.}(2007){Noeske}, {Weiner}, {Faber}, {Papovich},
  {Koo}, {Somerville}, {Bundy}, {Conselice}, {Newman}, {Schiminovich}, {Le
  Floc'h}, {Coil}, {Rieke}, {Lotz}, {Primack}, {Barmby}, {Cooper}, {Davis},
  {Ellis}, {Fazio}, {Guhathakurta}, {Huang}, {Kassin}, {Martin}, {Phillips},
  {Rich}, {Small}, {Willmer}, \& {Wilson}}]{noeske07}
{Noeske}, K.~G., {Weiner}, B.~J., {Faber}, S.~M., {et~al.} 2007, The
  Astrophysical Journal, 660, L43

\bibitem[{{Noll} {et~al.}(2009){Noll}, {Burgarella}, {Giovannoli}, {Buat},
  {Marcillac}, \& {Mu{\~n}oz-Mateos}}]{noll09}
{Noll}, S., {Burgarella}, D., {Giovannoli}, E., {et~al.} 2009, \aap, 507, 1793

\bibitem[{{Pacucci} \& {Loeb}(2024)}]{pacucci24}
{Pacucci}, F. \& {Loeb}, A. 2024, arXiv e-prints, arXiv:2401.04159

\bibitem[{{Perna} {et~al.}(2015{\natexlab{a}}){Perna}, {Brusa}, {Cresci},
  {Comastri}, {Lanzuisi}, {Lusso}, {Marconi}, {Salvato}, {Zamorani},
  {Bongiorno}, {Mainieri}, {Maiolino}, \& {Mignoli}}]{perna15a}
{Perna}, M., {Brusa}, M., {Cresci}, G., {et~al.} 2015{\natexlab{a}}, \aap, 574,
  A82

\bibitem[{{Perna} {et~al.}(2015{\natexlab{b}}){Perna}, {Brusa}, {Salvato},
  {Cresci}, {Lanzuisi}, {Berta}, {Delvecchio}, {Fiore}, {Lutz}, {Le Floc'h},
  {Mainieri}, \& {Riguccini}}]{perna15b}
{Perna}, M., {Brusa}, M., {Salvato}, M., {et~al.} 2015{\natexlab{b}}, \aap,
  583, A72

\bibitem[{{Pozzi} {et~al.}(2010){Pozzi}, {Vignali}, {Comastri}, {Bellocchi},
  {Fritz}, {Gruppioni}, {Mignoli}, {Maiolino}, {Pozzetti}, {Brusa}, {Fiore}, \&
  {Zamorani}}]{pozzi10}
{Pozzi}, F., {Vignali}, C., {Comastri}, A., {et~al.} 2010, \aap, 517, A11

\bibitem[{{Reines} \& {Volonteri}(2015)}]{reines15}
{Reines}, A.~E. \& {Volonteri}, M. 2015, \apj, 813, 82

\bibitem[{{Rosario} {et~al.}(2012){Rosario}, {Santini}, {Lutz}, {Shao},
  {Maiolino}, {Alexander}, {Altieri}, {Andreani}, {Aussel}, {Bauer}, {Berta},
  {Bongiovanni}, {Brandt}, {Brusa}, {Cepa}, {Cimatti}, {Cox}, {Daddi}, {Elbaz},
  {Fontana}, {F{\"o}rster Schreiber}, {Genzel}, {Grazian}, {Le Floch},
  {Magnelli}, {Mainieri}, {Netzer}, {Nordon}, {P{\'e}rez Garcia}, {Poglitsch},
  {Popesso}, {Pozzi}, {Riguccini}, {Rodighiero}, {Salvato}, {Sanchez-Portal},
  {Sturm}, {Tacconi}, {Valtchanov}, \& {Wuyts}}]{rosario12}
{Rosario}, D.~J., {Santini}, P., {Lutz}, D., {et~al.} 2012, \aap, 545, A45

\bibitem[{{Rosario} {et~al.}(2013){Rosario}, {Trakhtenbrot}, {Lutz}, {Netzer},
  {Trump}, {Silverman}, {Schramm}, {Lusso}, {Berta}, {Bongiorno}, {Brusa},
  {F{\"o}rster-Schreiber}, {Genzel}, {Lilly}, {Magnelli}, {Mainieri},
  {Maiolino}, {Merloni}, {Mignoli}, {Nordon}, {Popesso}, {Salvato}, {Santini},
  {Tacconi}, \& {Zamorani}}]{rosario13}
{Rosario}, D.~J., {Trakhtenbrot}, B., {Lutz}, D., {et~al.} 2013, \aap, 560, A72

\bibitem[{{Satyapal} {et~al.}(2014){Satyapal}, {Ellison}, {McAlpine}, {Hickox},
  {Patton}, \& {Mendel}}]{satyapal14}
{Satyapal}, S., {Ellison}, S.~L., {McAlpine}, W., {et~al.} 2014, \mnras, 441,
  1297

\bibitem[{{Schawinski} {et~al.}(2015){Schawinski}, {Koss}, {Berney}, \&
  {Sartori}}]{Schawinski15}
{Schawinski}, K., {Koss}, M., {Berney}, S., \& {Sartori}, L.~F. 2015, \mnras,
  451, 2517

\bibitem[{{Schreiber} {et~al.}(2015){Schreiber}, {Pannella}, {Elbaz},
  {B{\'e}thermin}, {Inami}, {Dickinson}, {Magnelli}, {Wang}, {Aussel}, {Daddi},
  {Juneau}, {Shu}, {Sargent}, {Buat}, {Faber}, {Ferguson}, {Giavalisco},
  {Koekemoer}, {Magdis}, {Morrison}, {Papovich}, {Santini}, \&
  {Scott}}]{schreiber15}
{Schreiber}, C., {Pannella}, M., {Elbaz}, D., {et~al.} 2015, Astronomy and
  Astrophysics, 575, A74

\bibitem[{{Scoville} {et~al.}(2007){Scoville}, {Aussel}, {Brusa}, {Capak},
  {Carollo}, {Elvis}, {Giavalisco}, {Guzzo}, {Hasinger}, {Impey}, {Kneib},
  {LeFevre}, {Lilly}, {Mobasher}, {Renzini}, {Rich}, {Sanders}, {Schinnerer},
  {Schminovich}, {Shopbell}, {Taniguchi}, \& {Tyson}}]{scoville07}
{Scoville}, N., {Aussel}, H., {Brusa}, M., {et~al.} 2007, \apjs, 172, 1

\bibitem[{{Secrest} {et~al.}(2020){Secrest}, {Ellison}, {Satyapal}, \&
  {Blecha}}]{secrest20}
{Secrest}, N.~J., {Ellison}, S.~L., {Satyapal}, S., \& {Blecha}, L. 2020,
  \mnras, 499, 2380

\bibitem[{{Shankar} {et~al.}(2016){Shankar}, {Bernardi}, {Sheth}, {Ferrarese},
  {Graham}, {Savorgnan}, {Allevato}, {Marconi}, {L{\"a}sker}, \&
  {Lapi}}]{shankar16}
{Shankar}, F., {Bernardi}, M., {Sheth}, R.~K., {et~al.} 2016, \mnras, 460, 3119

\bibitem[{{Silk} \& {Rees}(1998)}]{silk98}
{Silk}, J. \& {Rees}, M.~J. 1998, \aap, 331, L1

\bibitem[{{Spergel} {et~al.}(2003){Spergel}, {Verde}, {Peiris}, {Komatsu},
  {Nolta}, {Bennett}, {Halpern}, {Hinshaw}, {Jarosik}, {Kogut}, {Limon},
  {Meyer}, {Page}, {Tucker}, {Weiland}, {Wollack}, \& {Wright}}]{spergel03}
{Spergel}, D.~N., {Verde}, L., {Peiris}, H.~V., {et~al.} 2003, \apjs, 148, 175

\bibitem[{{Spinoglio} {et~al.}(2021){Spinoglio}, {Mordini},
  {Fern{\'a}ndez-Ontiveros}, {Alonso-Herrero}, {Armus}, {Bisigello}, {Calura},
  {Carrera}, {Cooray}, {Dannerbauer}, {Decarli}, {Egami}, {Elbaz},
  {Franceschini}, {Gonz{\'a}lez Alfonso}, {Graziani}, {Gruppioni},
  {Hatziminaoglou}, {Kaneda}, {Kohno}, {Labiano}, {Magdis}, {Malkan},
  {Matsuhara}, {Nagao}, {Naylor}, {Pereira-Santaella}, {Pozzi}, {Rodighiero},
  {Roelfsema}, {Serjeant}, {Vignali}, {Wang}, \& {Yamada}}]{spinoglio21}
{Spinoglio}, L., {Mordini}, S., {Fern{\'a}ndez-Ontiveros}, J.~A., {et~al.}
  2021, \pasa, 38, e021

\bibitem[{{Stalevski} {et~al.}(2012){Stalevski}, {Fritz}, {Baes}, {Nakos}, \&
  {Popovic}}]{stalevski12}
{Stalevski}, M., {Fritz}, J., {Baes}, M., {Nakos}, T., \& {Popovic}, L.~C.
  2012, Publications de l'Observatoire Astronomique de Beograd, 91, 235

\bibitem[{{Stalevski} {et~al.}(2016){Stalevski}, {Ricci}, {Ueda}, {Lira},
  {Fritz}, \& {Baes}}]{stalevski16}
{Stalevski}, M., {Ricci}, C., {Ueda}, Y., {et~al.} 2016, \mnras, 458, 2288

\bibitem[{{Stanley} {et~al.}(2015){Stanley}, {Harrison}, {Alexander},
  {Swinbank}, {Aird}, {Del Moro}, {Hickox}, \& {Mullaney}}]{stanley15}
{Stanley}, F., {Harrison}, C.~M., {Alexander}, D.~M., {et~al.} 2015, \mnras,
  453, 591

\bibitem[{{Stemo} {et~al.}(2020){Stemo}, {Comerford}, {Barrows}, {Stern},
  {Assef}, \& {Griffith}}]{stemo20}
{Stemo}, A., {Comerford}, J.~M., {Barrows}, R.~S., {et~al.} 2020, \apj, 888, 78

\bibitem[{{Suh} {et~al.}(2020){Suh}, {Civano}, {Trakhtenbrot}, {Shankar},
  {Hasinger}, {Sanders}, \& {Allevato}}]{suh20}
{Suh}, H., {Civano}, F., {Trakhtenbrot}, B., {et~al.} 2020, \apj, 889, 32

\bibitem[{{Tacconi} {et~al.}(2020){Tacconi}, {Genzel}, \&
  {Sternberg}}]{tacconi20}
{Tacconi}, L.~J., {Genzel}, R., \& {Sternberg}, A. 2020, \araa, 58, 157

\bibitem[{{Toba} {et~al.}(2017){Toba}, {Bae}, {Nagao}, {Woo}, {Wang}, {Wagner},
  {Sun}, \& {Chang}}]{toba17}
{Toba}, Y., {Bae}, H.-J., {Nagao}, T., {et~al.} 2017, \apj, 850, 140

\bibitem[{{Treister} {et~al.}(2012){Treister}, {Schawinski}, {Urry}, \&
  {Simmons}}]{treister12}
{Treister}, E., {Schawinski}, K., {Urry}, C.~M., \& {Simmons}, B.~D. 2012,
  \apjl, 758, L39

\bibitem[{{{\"U}bler} {et~al.}(2023){{\"U}bler}, {Maiolino}, {Curtis-Lake},
  {P{\'e}rez-Gonz{\'a}lez}, {Curti}, {Perna}, {Arribas}, {Charlot}, {Marshall},
  {D'Eugenio}, {Scholtz}, {Bunker}, {Carniani}, {Ferruit}, {Jakobsen}, {Rix},
  {Rodr{\'\i}guez Del Pino}, {Willott}, {Boeker}, {Cresci}, {Jones}, {Kumari},
  \& {Rawle}}]{ubler23}
{{\"U}bler}, H., {Maiolino}, R., {Curtis-Lake}, E., {et~al.} 2023, \aap, 677,
  A145

\bibitem[{{Ueda} {et~al.}(2014){Ueda}, {Akiyama}, {Hasinger}, {Miyaji}, \&
  {Watson}}]{ueda14}
{Ueda}, Y., {Akiyama}, M., {Hasinger}, G., {Miyaji}, T., \& {Watson}, M.~G.
  2014, \apj, 786, 104

\bibitem[{{Vignali}(2014)}]{vignali14_osc}
{Vignali}, C. 2014, in IAU Symposium, Vol. 304, Multiwavelength AGN Surveys and
  Studies, ed. A.~M. {Mickaelian} \& D.~B. {Sanders}, 132--138

\bibitem[{{Vignali} {et~al.}(2006){Vignali}, {Alexander}, \&
  {Comastri}}]{vignali06}
{Vignali}, C., {Alexander}, D.~M., \& {Comastri}, A. 2006, \mnras, 373, 321

\bibitem[{{Vignali} {et~al.}(2010){Vignali}, {Alexander}, {Gilli}, \&
  {Pozzi}}]{vignali10}
{Vignali}, C., {Alexander}, D.~M., {Gilli}, R., \& {Pozzi}, F. 2010, \mnras,
  404, 48

\bibitem[{{Vignali} {et~al.}(2014){Vignali}, {Mignoli}, {Gilli}, {Comastri},
  {Iwasawa}, {Zamorani}, {Mainieri}, \& {Bongiorno}}]{vignali14}
{Vignali}, C., {Mignoli}, M., {Gilli}, R., {et~al.} 2014, \aap, 571, A34

\bibitem[{{Weaver} {et~al.}(2022){Weaver}, {Kauffmann}, {Ilbert}, {McCracken},
  {Moneti}, {Toft}, {Brammer}, {Shuntov}, {Davidzon}, {Hsieh}, {Laigle},
  {Anastasiou}, {Jespersen}, {Vinther}, {Capak}, {Casey}, {McPartland},
  {Milvang-Jensen}, {Mobasher}, {Sanders}, {Zalesky}, {Arnouts}, {Aussel},
  {Dunlop}, {Faisst}, {Franx}, {Furtak}, {Fynbo}, {Gould}, {Greve}, {Gwyn},
  {Kartaltepe}, {Kashino}, {Koekemoer}, {Kokorev}, {Le F{\`e}vre}, {Lilly},
  {Masters}, {Magdis}, {Mehta}, {Peng}, {Riechers}, {Salvato}, {Sawicki},
  {Scarlata}, {Scoville}, {Shirley}, {Silverman}, {Sneppen}, {Smolc̆i{\'c}},
  {Steinhardt}, {Stern}, {Tanaka}, {Taniguchi}, {Teplitz}, {Vaccari}, {Wang},
  \& {Zamorani}}]{cosmos2020}
{Weaver}, J.~R., {Kauffmann}, O.~B., {Ilbert}, O., {et~al.} 2022, \apjs, 258,
  11

\bibitem[{{Xue} {et~al.}(2011){Xue}, {Luo}, {Brandt}, {Bauer}, {Lehmer},
  {Broos}, {Schneider}, {Alexand er}, {Brusa}, {Comastri}, {Fabian}, {Gilli},
  {Hasinger}, {Hornschemeier}, {Koekemoer}, {Liu}, {Mainieri}, {Paolillo},
  {Rafferty}, {Rosati}, {Shemmer}, {Silverman}, {Smail}, {Tozzi}, \&
  {Vignali}}]{xue11}
{Xue}, Y.~Q., {Luo}, B., {Brandt}, W.~N., {et~al.} 2011, \apjs, 195, 10

\bibitem[{{Yang} {et~al.}(2022){Yang}, {Boquien}, {Brandt}, {Buat},
  {Burgarella}, {Ciesla}, {Lehmer}, {Ma{\l}ek}, {Mountrichas}, {Papovich},
  {Pons}, {Stalevski}, {Theul{\'e}}, \& {Zhu}}]{yang22}
{Yang}, G., {Boquien}, M., {Brandt}, W.~N., {et~al.} 2022, \apj, 927, 192

\bibitem[{{Yang} {et~al.}(2020){Yang}, {Boquien}, {Buat}, {Burgarella},
  {Ciesla}, {Duras}, {Stalevski}, {Brandt}, \& {Papovich}}]{yang20}
{Yang}, G., {Boquien}, M., {Buat}, V., {et~al.} 2020, \mnras, 491, 740

\bibitem[{{Zakamska} {et~al.}(2016){Zakamska}, {Hamann}, {P{\^a}ris}, {Brandt},
  {Greene}, {Strauss}, {Villforth}, {Wylezalek}, {Alexandroff}, \&
  {Ross}}]{zakamska16}
{Zakamska}, N.~L., {Hamann}, F., {P{\^a}ris}, I., {et~al.} 2016, \mnras, 459,
  3144

\bibitem[{{Zhao} {et~al.}(2021){Zhao}, {Marchesi}, {Ajello}, {Cole}, {Hu},
  {Silver}, \& {Torres-Alb{\`a}}}]{zhao21}
{Zhao}, X., {Marchesi}, S., {Ajello}, M., {et~al.} 2021, \aap, 650, A57

\end{thebibliography}

\begin{appendix}

\section{X-ray analysis}\label{sec:xray_prop}
We performed the spectral fitting via the \texttt{XSPEC} package \citep{xspec} with two different models. The first consists of a power-law component (\texttt{powerlaw}) modified by the Galactic absorption (\texttt{phabs}) at the source position, computed via the \texttt{nh} tool, which derives it from the HI map by \citet{kalberla05}. The second model includes a power-law component with fixed photon index (representing the intrinsic un-absorbed emission of the AGN), modified by the AGN torus obscuration via a photoelectric absorption component ($N_H$) at the source redshift (\texttt{zphabs}), and by the Galactic absorption. The photon index had to be fixed to $\Gamma=1.8$, a typical value for the intrinsic AGN emission, due to the degeneracy between $\Gamma$ and $N_H$ that cannot be resolved in most of our spectra due to the low number of net-counts: the flat X-ray spectra of an obscured AGN can be fitted equally well with a steep power-law and significant absorption as with a flat power-law and very low absorption.  In Table \ref{tab:xray_prop}, we report the spectral properties of all the X-ray- detected sources.

\begin{table}[h]
\caption{Properties of the X-ray-detected sample obtained from an X-ray spectral analysis.  }
\label{tab:xray_prop}
\centering
\resizebox{\hsize}{!}{
\begin{tabular}{ccccccc}
\hline \hline
ID & $z$ & Ncts & $\Gamma$ & $\log{(N_{H}/ \rm{cm^{-2}})}$ & $\log{(L_{2-10keV}/\rm{L_{\odot}})}$ & X/\nev\ \\ 
\hline
lid1856 & 0.73 & 1827 & $1.6_{-0.1}^{+0.1}$ & $21.6_{-0.2}^{+0.1}$ & $44.41_{-0.03}^{+0.02}$ & $1456_{-86}^{+147}$\\[3pt]
cid339 & 0.69 & 373 & $1.5_{-0.2}^{+0.2}$ & $21.7_{-0.4}^{+0.2}$ & $43.61_{-0.06}^{+0.06}$ & $604_{-84}^{+81}$\\[3pt]
cid522 & 0.94 & 336 & $1.1_{-0.2}^{+0.2}$ & $22.5_{-0.1}^{+0.1}$ & $44.10_{-0.05}^{+0.06}$ & $531_{-64}^{+62}$\\[3pt]
cid110 & 0.73 & 313 & $1.6_{-0.2}^{+0.2}$ & $<21.7$ & $43.56_{-0.06}^{+0.06}$ & $667_{-167}^{+163}$\\[3pt]
cid173 & 1.00 & 242 & $1.5_{-0.2}^{+0.3}$ & $22.0_{-1.3}^{+0.3}$ & $43.99_{-0.08}^{+0.07}$ & $1182_{-150}^{+150}$\\[3pt]
lid1840 & 1.01 & 216 & $2.0_{-0.3}^{+0.3}$ & $<21.7$ & $43.71_{-0.06}^{+0.05}$ & $360_{-66}^{+46}$\\[3pt]
cid381 & 0.86 & 208 & $0.9_{-0.3}^{+0.2}$ & $22.8_{-0.2}^{+0.2}$ & $43.8_{-0.1}^{+0.1}$ & $284_{-41}^{+38}$\\[3pt]
lid279 & 0.88 & 173 & $1.9_{-0.3}^{+0.4}$ & $<21.7$ & $43.46_{-0.07}^{+0.06}$ & $810_{-157}^{+147}$\\[3pt]
lid1478 & 0.83 & 168 & $1.0_{-0.3}^{+0.3}$ & $22.6_{-0.3}^{+0.2}$ & $43.7_{-0.1}^{+0.1}$ & $332_{-57}^{+54}$\\[3pt]
cid496 & 0.90 & 165 & $1.5_{-0.3}^{+0.3}$ & $22.1_{-1.1}^{+0.3}$ & $43.6_{-0.1}^{+0.1}$ & $98_{-17}^{+15}$\\[3pt]
cid456 & 1.02 & 162 & $0.2_{-0.4}^{+0.3}$ & $23.3_{-0.2}^{+0.2}$ & $44.2_{-0.1}^{+0.2}$ & $113_{-18}^{+18}$\\[3pt]
cid221 & 0.75 & 142 & $1.8_{-0.3}^{+0.3}$ & $<22$ & $43.3_{-0.1}^{+0.1}$ & $461_{-102}^{+96}$\\[3pt]
cid620 & 1.18 & 126 & $0.7_{-0.4}^{+0.4}$ & $23.4_{-0.2}^{+0.2}$ & $44.3_{-0.1}^{+0.1}$ & $146_{-25}^{+27}$\\[3pt]
lid1826 & 1.17 & 109 & $1.2_{-0.3}^{+0.3}$ & $22.9_{-0.3}^{+0.2}$ & $44.0_{-0.1}^{+0.1}$ & $122_{-24}^{+26}$\\[3pt]
cid138 & 0.70 & 99 & $1.4_{-0.4}^{+0.4}$ & $22.3_{-1.0}^{+0.3}$ & $43.2_{-0.2}^{+0.1}$ & $439_{-108}^{+94}$\\[3pt]
cid1126 & 0.96 & 93 & $0.2_{-0.6}^{+0.5}$ & $23.2_{-0.2}^{+0.2}$ & $43.9_{-0.2}^{+0.2}$ & $186_{-58}^{+58}$\\[3pt]
lid489 & 0.85 & 93 & $1.7_{-0.5}^{+0.5}$ & $<22.1$ & $43.8_{-0.1}^{+0.1}$ & $643_{-263}^{+224}$\\[3pt]
cid717 & 0.89 & 85 & $1.5_{-0.3}^{+0.3}$ & $>21.0$ & $43.2_{-0.1}^{+0.1}$ & $122_{-26}^{+26}$\\[3pt]
cid503 & 0.91 & 80 & $0.3_{-0.4}^{+0.4}$ & $22.8_{-0.2}^{+0.2}$ & $43.6_{-0.1}^{+0.1}$ & $566_{-122}^{+122}$\\[3pt]
cid426 & 0.86 & 75 & $0.5_{-0.4}^{+0.4}$ & $22.7_{-0.1}^{+0.2}$ & $43.5_{-0.1}^{+0.1}$ & $290_{-66}^{+61}$\\[3pt]
cid254 & 0.71 & 60 & $1.2_{-0.4}^{+0.4}$ & $22.1_{-0.5}^{+0.3}$ & $43.0_{-0.1}^{+0.1}$ & $143_{-55}^{+52}$\\[3pt]
lid689 & 0.68 & 59 & $2.1_{-0.4}^{+0.4}$ & $<21.6$ & $43.09_{-0.10}^{+0.09}$ & $253_{-71}^{+59}$\\[3pt]
lid3483 & 0.66 & 51 & $-0.4_{-0.6}^{+0.5}$ & $23.3_{-0.2}^{+0.2}$ & $43.5_{-0.1}^{+0.1}$ & $310_{-89}^{+80}$\\[3pt]
cid1130 & 0.79 & 48 & $1.4_{-0.5}^{+0.5}$ & $>21.9$ & $42.9_{-0.1}^{+0.1}$ & $152_{-52}^{+47}$\\[3pt]
cid1019 & 0.73 & 41 & $-1.0_{-1.7}^{+1.9}$ & $24.2_{-0.6}^{+0.3}$ & $44.1_{-1.0}^{+1.1}$ & <105\\[3pt]
lid1603 & 0.97 & 39 & $1.1_{-0.5}^{+0.6}$ & $>22.7$ & $43.1_{-0.2}^{+0.2}$ & $110_{-39}^{+35}$\\[3pt]
cid401 & 0.97 & 39 & $-0.4_{-0.6}^{+0.6}$ & $23.3_{-0.2}^{+0.3}$ & $43.6_{-0.2}^{+0.2}$ & $129_{-49}^{+41}$\\[3pt]
lid1869 & 1.17 & 36 & $0.3_{-0.6}^{+0.6}$ & $23.0_{-0.3}^{+0.3}$ & $43.6_{-0.2}^{+0.2}$ & $93_{-26}^{+24}$\\[3pt]
cid1230 & 0.75 & 35 & $1.4_{-0.6}^{+0.6}$ & $>22.0$ & $42.6_{-0.1}^{+0.2}$ & $40_{-15}^{+13}$\\[3pt]
cid1169 & 0.96 & 21 & $0.9_{-0.8}^{+0.9}$ & $22.6_{-1.0}^{+0.4}$ & $42.9_{-0.3}^{+0.3}$ & $167_{-90}^{+77}$\\[3pt]
lid2210 & 0.74 & 18 & $0.8_{-0.9}^{+0.8}$ & $22.6_{-0.4}^{+0.3}$ & $42.7_{-0.3}^{+0.3}$ & $43_{-28}^{+22}$\\[3pt]
lid1459 & 1.00 & 17 & $1.0_{-1.0}^{+1.0}$ & $22.8_{-0.3}^{+0.5}$ & $43.3_{-0.3}^{+0.3}$ & $46_{-24}^{+17}$\\[3pt]
cid1706 & 0.76 & 15 & $-1.6_{-1.0}^{+1.3}$ & $24.0_{-0.4}^{+0.3}$ & $43.9_{-0.6}^{+0.6}$ & <107\\[3pt]
cid2454 & 0.76 & 15 & $-0.1_{-2.1}^{+2.2}$ & $23.3_{-1.0}^{+0.7}$ & $43.2_{-0.4}^{+0.3}$ & $84_{-81}^{+57}$\\[3pt]
lid3017 & 0.68 & 15 & $-0.1_{-1.2}^{+0.9}$ & $22.9_{-0.5}^{+1.1}$ & $42.7_{-0.4}^{+1.0}$ & $60_{-35}^{+30}$\\[3pt]
cid1508 & 0.67 & 11 & $-0.9_{-0.9}^{+1.5}$ & $24.0_{-1.1}^{+0.6}$ & $41.9_{-0.7}^{+0.5}$ & <89\\[3pt]
\hline
\end{tabular}}
\tablefoot{
\textit{Ncts} refers to the number of net-counts (i.e., background subtracted), $\Gamma$ to the photon index obtained using a power-law model with Galactic absorption. The amount of obscuration ($N_{H}$) and intrinsic (i.e. absorption corrected) $2-10$ keV luminosity are obtained from a model with Galactic absorption, power-law emission with fixed photon index ($\Gamma=1.8$), and absorption component at the source redshift to model the source obscuration.}
\end{table}

\section{Torus models}\label{sec:torusmodels}
The torus library we used to model the AGN contribution to the SED assumes that the AGN dust and gas are distributed in a toroidal shape, namely, the ``smooth-torus'' model. It was developed by \citet{fritz06} and updated by \citet{feltre12}. The geometry of the torus is that of a "flared" disc. Its size is defined by the outer radius $R_{\text{max}}$ -- the inner radius being defined by the sublimation temperature of dust grains under the influence of the strong nuclear radiation field - and by the angular opening angle $\Theta$ of the torus itself.
The main dust components are silicate and graphite grains, in almost equal percentages. The torus density law adopted is:
\begin{equation}
\label{eq:torus_density}
\rho(r,\theta)=\alpha \times r^{\beta} \times e^{-\gamma\, |\cos{\theta}|}
,\end{equation}
where $\alpha$ is a normalization constant and the parameters $\beta$ and $\gamma$ allow us to create density gradients both in radial ($r$) and in polar ($\theta$) directions.
The models assume that the torus is illuminated by a central point-like energy source with isotropic emission. Its spectrum is described as a composition of power-laws with variable indices (see \citealt{feltre12}). The radiation emitted is given by the sum of the primary source located in the torus center and a secondary contribution given by thermal and scattering dust emission.\par
To reduce the calculation time, we selected only a sub-sample of the $24\,000$ elements torus library. We chose:
\begin{itemize}
\item{$\mathbf{\Phi=1^{\circ}, 21^{\circ}, 41^{\circ}, 61^{\circ}, 89^{\circ}}$:} to be able to model different inclination angles between the line of sight and the torus equatorial plane (i.e., to model both type 1 and type 2 objects).
\item{$\mathbf{R=30\,:}$} this value limits the models to compact tori of a few parsec (given that $R_{\text{min}}$ is directly connected to the sublimation temperature and to the accretion luminosity of the central BH), as done in \citet{pozzi10}. In fact, high-resolution IR and recent ALMA observations support a compact dust distribution in nearby luminous AGNs (i.e., \citep{jaffe04,elitzur08,combes19}).
\item{$\mathbf{ct=20^{\circ}, 40^{\circ}, 60^{\circ}}$:} all the possible values of the half-width of the torus apertures.
\item{$\mathbf{\beta=0, -1}$:} the first is linked to an homogeneous density distribution, the second to a density decreasing exponentially with the distance from the nucleus. 
\item{$\mathbf{\gamma=0}$:} we considered only torus with an homogeneous distribution of density in polar direction.
\item{$\mathbf{\tau_{\text{eq}}=0.1, 0.3, 0.6, 1, 3, 6}$:} as suggested by \citet{feltre12}, we avoided extreme optical depths.
\end{itemize}
Reducing the torus parameter space allowed us to reduce the calculation time to an acceptable level while maintaining 180 different torus models. We instructed the SED fitting algorithm to run 100 normalizations for each torus model, for a total of $18\,000$ torus SEDs available. 

\section{$M_{\text{BH}}-M_*$ relations}\label{sec:mbhrelations}

\begin{figure}[h]
  \centering
  \resizebox{\hsize}{!}{\includegraphics{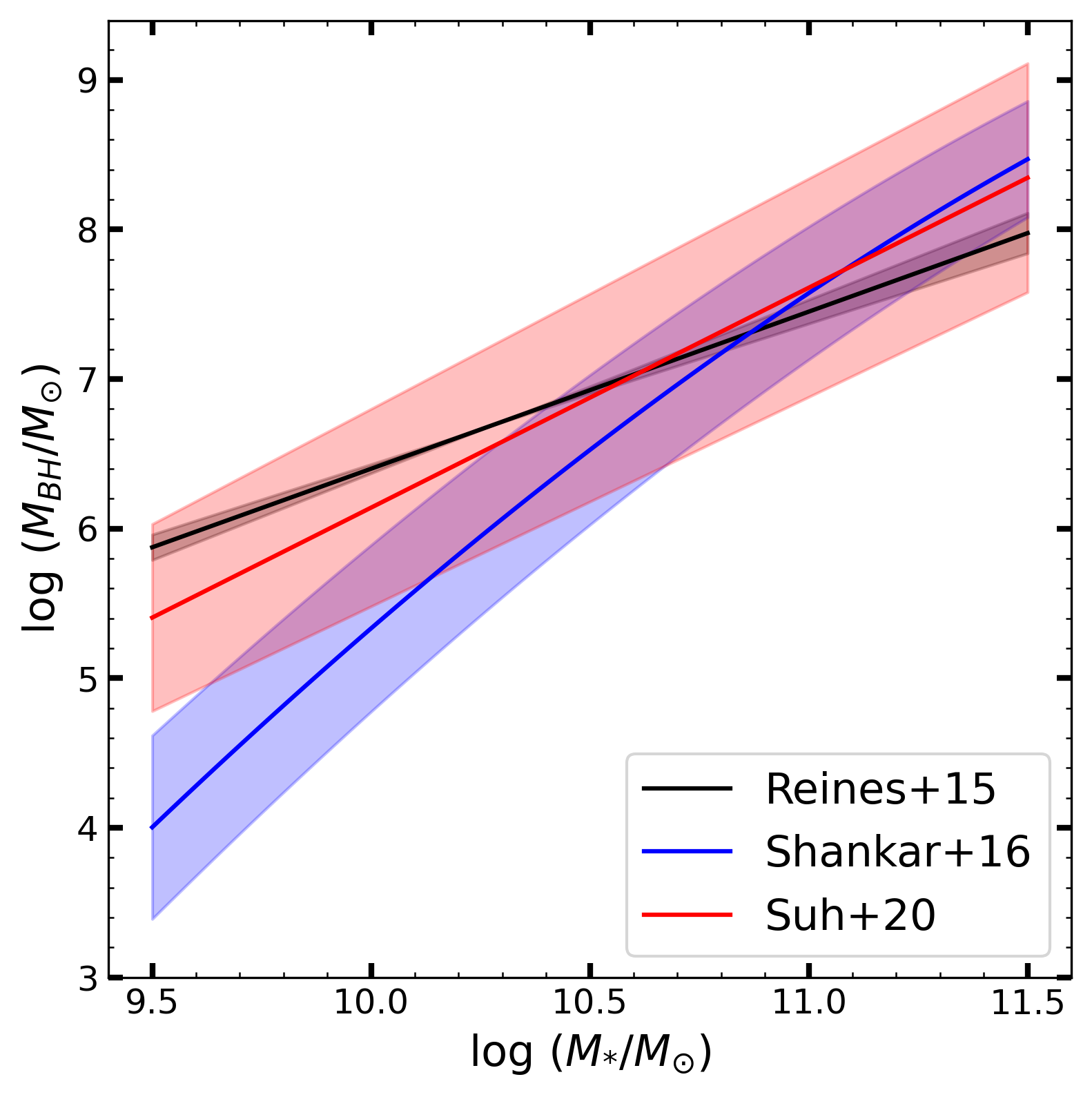}}
  \caption{Comparison of different $M_{\text{BH}}-M_*$ relations and relative uncertainties. The red line refers to the \citet{suh20} relation used in this work, while the black, and blue to the \citet{reines15}, and \citet{shankar16} respectively. Using the \citet{reines15}, or \citet{shankar16}, does not significantly impact  our results.}
  \label{fig:mbh_mstar}
\end{figure}

Due to the high number of different $M_{\text{BH}}-M_*$ relations in literature, their large uncertainties and the different ways used to estimate the \mbh\, we tested three different M$_{\text{BH}} - \text{M}_{*}$ relations and compared them in Fig.\ref{fig:mbh_mstar}.\par
The \citet{suh20} $M_{\text{BH}} - \text{M}_{*}$ relation (red line) was obtained from a sample of 100 X-ray selected AGNs with the BH masses computed considering single epoch H$\alpha$, H$\beta$, and \mgii\ broad line widths and line/continuum luminosity as proxy for the size and velocity of the BBLR. Thus, their sample was composed of active galaxies but only of type 1 AGNs.
The \citet{reines15} relation (black line) comes from a sample of 262 broad-line AGNs and 79 galaxies. For the AGNs, their \mbh estimations are derived from single epoch spectra of sources with broad H$\alpha$, using the line FWHM as well as its luminosity, under the virial assumption. For 15 AGNs, they used reverberation-mapped \mbh\ from literature. Finally, the \mbh for the 79 galaxies were obtained from measurements based on stellar dynamics, gas dynamics, and maser disk dynamics. 
The blue line is the \citet{shankar16} relation, obtained with sources from five different literature samples of galaxies with BH dynamical mass measurements.\par
For our work, we chose the \citet{suh20},$M_{\text{BH}} - \text{M}_{*}$ relation to estimating the \mbh\ mass for the following considerations: as our sample is composed of AGNs, we preferred to avoid relations derived only with non-active galaxies; in addition, the \citet{suh20} relation is a ``middle ground'' compromise between the three relations and, considering the uncertainties, is compatible with the other relations. Finally, we note that in the $10\geq \log{(\rm{M_*}/M_\odot)} \geq11.2$ mass range where most of our sources are, the scatters in the relationship overlap significantly.

\end{appendix}

\end{document}